\documentclass[sn-basic]{sn-jnl}

\usepackage{graphicx}%
\usepackage{multirow}%
\usepackage{amsmath,amssymb,amsfonts}%
\usepackage{amsthm}%
\usepackage{mathrsfs}%
\usepackage[title]{appendix}%
\usepackage{xcolor}%
\usepackage{textcomp}%
\usepackage{manyfoot}%
\usepackage{booktabs}%
\usepackage{algorithm}%
\usepackage{algorithmicx}%
\usepackage{algpseudocode}%
\usepackage{listings}%

 \pdfoutput=1

 \newcommand\nc\newcommand
\nc\vect[1]{\boldsymbol{#1}}
\nc\be{\begin{equation}}
\nc\ee{\end{equation}}
\newcommand\bleu[1]{\bf\color{blue}{#1}} 
\newcommand\rouge[1]{\bf\color{red}{#1}} 

\usepackage{changes}

\begin{document}

\title[High-Frequency Radar observation of strong and contrasted currents: the Alderney race paradigm]{High-Frequency Radar observation of strong and sheared currents: the Alderney race paradigm}


\author[1]{\fnm{Dylan} \sur{Dumas}}\email{dylan.dumas@mio.osupytheas.fr}
\author[2]{\fnm{Anne-Claire} \sur{Bennis}}\email{anne-claire.bennis@unicaen.fr}
\author*[1]{\fnm{Charles-Antoine} \sur{Gu{\'e}rin}}\email{guerin@univ-tln.fr}
\author[3]{\fnm{Guiomar} \sur{Lopez}}\email{glopez@socib.es}
\author[2]{\fnm{Laurent} \sur{Benoit}}\email{{laurent.benoit@unicaen.fr}}


\affil*[1]{\orgdiv{MIO}, \orgname{Univ Toulon, CNRS, AMU, IRD}, \orgaddress{\city{Toulon},\country{France}}}
\affil[2]{{\orgdiv{M2C}, \orgname{Normandie Univ., UNICAEN, UNIROUEN, CNRS}, \orgaddress{\city{Caen}, \country{France}}}}
\affil[3]{\orgdiv{SOCIB}, \orgname{{Balearic Islands Coastal Observing and Forecasting System}, \orgaddress{\city{Palma}, \country{Spain}}}}


\abstract{The Alderney Race has been identified as a future site for the development of tidal energy, due to its bidirectional strong current reaching 5 m/s during spring tides. This hydrodynamics is very difficult to measure by {\textit{in situ}} or remote sensing means. High-frequency coastal radars can provide a synoptic and near-real-time view of such a complex circulation, but the classical processing algorithms are not adapted to the extreme situation of strongly sheared currents. We propose an improved high-resolution direction-finding technique for the azimuthal processing of such radar data. It uses phased-array systems and combines the advantages of the usual beam-forming technique to eliminate many problems related to the distortion of Doppler spectra by extreme currents. The method is evaluated with a unique data set of radar measurements at two radar frequencies (13 and 24.5 MHz) and three spatial resolutions (200, 750, and 1500 m). The radar-based surface currents are analyzed in the light of a high-resolution numerical model and also compared with {\textit{in situ}} measurements. While high azimuthal resolution can be achieved in this way, it is shown that the typical range resolutions of 750 and 1500 m are insufficient to account for the strong spatial variations of the surface current at some specific times and locations.}




\maketitle

\section{Introduction}\label{sec1}

Changes in the Earth's climate due to anthropogenic effects are increasingly present {and} becoming more frequent. To mitigate these changes, some groups of countries, such as the EU, have agreed on the common goal of carbon neutrality. The development of marine renewable energy {including tidal energy} is a promising way to achieve this goal. Researchers have studied the complex hydrodynamics of the Alderney Race (France) {since} it is the most promising tidal site in Western Europe, with a maximum potential of $5.1 \hskip2pt \mathrm{GW}$ due to extreme tidal currents reaching $5 \hskip2pt \mathrm{m.s^{-1}}$ {\citep[e.g.][]{BaillyDuBois&al2012}}.

Instrumentation of the Alderney Race with classical \textit{in situ} instruments such as current meters, wave buoys, or pressure gauges is a challenging task due to the high exposure of these devices to damage and loss. In addition to this maintenance issue, another limitation is the punctual nature of the measurements, which is useful but not sufficient for the tidal industry.  An alternative is the use of land-based High-Frequency radar (HFR), which can provide a synoptic view of surface currents and sea state.  As is well known since the pioneering experiment of \cite{crombie_doppler_1955}, the dominant electromagnetic sea echo measured by HFR is due to the so-called Bragg wave, whose wavelength is half the radar wavelength. As a result, the backscattered ocean Doppler spectrum has two main peaks, called Bragg lines, at a predictable frequency, and the HFR radial surface current velocity is obtained by subtracting the observed position of the Bragg line from its theoretical position {\citep{Stewart&Joy1974}}. HFR can thus provide near real-time measurements of the surface layer current over a region typically extending a few tens of kilometers from shore, with a spatial resolution on the order of a few $\mathrm{km^{2}}$, depending on the operating frequency. The physics and technology behind HFR have reached a mature state. (see e.g. \cite{roarty_global_2019,lorente_OceanScience22,reyes_OceanScience22} for recent reviews) and these instruments give satisfactory estimates, with known limitations and accuracy, of the surface currents in the vast majority of situations. However, the non-standard case of very intense currents undergoing strong horizontal and vertical shears such as those found in the Alderney Race falls outside of the applicability domain of standard radar signal processing methods and requires dedicated treatments.
 
\cite{lopez2018,Lopez_PhilTrans20} analyzed the {h}ydrodynamics of the Alderney race using the first HFR datasets from two radar stations implemented at the Cape of La Hague in 2017, operating at $13.5\hskip2pt \mathrm{MHz}$ and $24.5\hskip2pt \mathrm{MHz}$ (more details on the HFR facilities are given in section \ref{sec2b}). The HFR surface velocity was compared with the surface velocity computed by a three-dimensional fully coupled wave-current model. Very good results were obtained, with RMSE (Root Mean Square Error) less than $0.4 \hskip2pt \mathrm{m.s^{-1}}$ and $0.6 \hskip2pt \mathrm{m.s^{-1}}$ during spring and neap tides, respectively.  The correlation coefficients are greater {than} $0.95$ over the entire HF region.  They found maximum errors in shallow areas with high mean current values ($1.25 \hskip2pt \mathrm{m.s^{-1}}$ and $2.7 \hskip2pt \mathrm{m.s^{-1}}$ during neap and spring tides, respectively).

In the literature, HF measurements and processing algorithms for strong currents (i.e., with a mean velocity above, say, $2 \hskip2pt \mathrm{m.s^{-1}}$) are rarely documented and, to the best of our knowledge, are limited to studies in the Bali Strait (Indonesia) and the Fromveur Strait (France). In the case of the Bali Strait, the tidal velocity can reach a maximum of $2.97 \hskip2pt \mathrm{m.s^{-1}}$ and $2.17 \hskip2pt \mathrm{m.s^{-1}}$ during ebb and flood, respectively. Data from two HFRs operating at $26.275\hskip2pt \mathrm{MHz}$ were tested against ADCP (Acoustic Doppler Current Profiler) data from {\cite{Firdaus&al2021}}. Good agreement was found between the two datasets with a correlation coefficient of $0.813$ and an RMSE of $0.22 \hskip2pt \mathrm{m.s^{-1}}$ for the meridional velocity.  In contrast, the HFR zonal velocity was found to be very different from the ADCP zonal velocity, with a correlation coefficient of $0.235$ and an RMSE of $0.402 \hskip2pt \mathrm{m.s^{-1}}$. The authors attributed this inaccuracy to the smaller magnitude of the zonal component and the unfavorable configuration of the HFR stations. Surface current monitoring in the Fromveur Strait (France) using a pair of HFRs has revealed a maximum tidal current of about $3.8 \hskip2pt \mathrm{m.s^{-1}}$ \cite{sentchev2013}. {\cite{Thiebaut&Sentchev2015,Thiebaut&Sentchev2017}} found a correlation coefficient of about $0.82$ and a zero phase lag between the HFR surface velocity and the ADCP depth-averaged velocity. They concluded that the HFR surface velocity is $15-20\%$ higher than the ADCP depth-averaged velocity, which is in agreement with \cite{prandle1993}. {In contrast, Fig. 6 and Fig. 10 of {\cite{Thiebaut&Sentchev2015,Thiebaut&Sentchev2017}} show a significant underestimation (by more than $1 \hskip2pt \mathrm{m.s^{-1}}$) of the absolute velocity, as shown in Fig. 4 of \cite{Lopez_PhilTrans20} for the Alderney race. 
The conflicting results of these studies suggest that the measurement of very strong and rapidly changing currents with HFR is not yet fully reliable and requires some improvements in the processing algorithms. They also raise fundamental questions about the relevant spatial and temporal resolution needed to observe such \textcolor{black}{sheared} currents.

In this study, we present radar processing methods adapted to the case of intense surface currents, whose velocity can be even higher than the Bragg wave velocity. This situation requires high-resolution azimuthal processing, which can only be achieved by Beam Forming (BF) processing using a large phased-array (16 antennas) or by high-performance Direction Finding (DF) processing. However, when strong currents are combined with a strong sea state, the latter approach fails due to the difficulty of identifying the first-order Bragg region in the omnidirectional Doppler spectra. We propose an adaptation of the DF method to solve this problem. For this purpose, we resort to several techniques recently proposed for phased-array radars {in} \cite{dumas_JTECH23} and complement them with a preconditioning of the first-order Bragg range with a preliminary BF. This new method, which we call "hybrid BF-DF", is applied to the HF data of the Alderney Race, and its performance is evaluated using the results of a three-dimensional numerical model and a comparison with a classical BF processing with a large number of receiving antennas.

While previous studies using the same HFR stations \citep{lopez2018,Lopez_PhilTrans20} dealt only with typical range resolutions of 1500 m and 750 m at 13 and 24.5 MHz, respectively, in this study we present an original data set acquired during one week at the very high spatial resolution of 200 m for the higher transmit frequency (24.5 MHz). The addition of this exceptional data set allows us to evaluate the impact of spatial resolution on the detection of first-order Bragg peaks and surface current mapping. We show that at some specific locations and times of the tidal cycle, a very fine spatial resolution is required to account for horizontal current shear, which can be misleading when averaged over a coarser radar cell. 

The paper is organized as follows. The study site and the HFR facilities are described in section \ref{sec2}. The radar signal processing methods and the improved algorithms are presented in section \ref{sec3}. Various results and performances of the HFR measurements in the light of numerical model outputs and {\textit{in situ}} measurements are shown and discussed in Section \ref{sec4}, together with a discussion of the influence of spatial resolution. Conclusions are presented in section \ref{sec5}.
\section{Site description}\label{sec2}

\subsection{The Alderney race}
\textcolor{black}{The study site, Alderney Race, is located in the English Channel, which separates France from the United Kingdom. The English Channel is an epicontinental sea with a maximum depth of about $200 \hskip2pt \mathrm{m}$ {\citep{Dauvin2019}}. The hydrodynamics are mainly driven by semi-diurnal tides, which produce two ebb and two flood tides each day, with a maximum tidal range of $14\hskip2pt \mathrm{m}$ in the {Mont-Saint-Michel Bay} during very large spring tides {\citep{Cayocca&al2008}}. Surface waves and wind also influence the hydrodynamics of the English Channel, especially during storms and near the coast {\citep[e.g.][]{Korotenko&al2012}}. The tidal current in the English Channel can also modify the surface winds and cause tidal winds {\citep{Renault&Marchesiello2022}}. }
The Alderney Race is located in the western part of the English Channel near the Cotentin Peninsula between the Cape of La Hague and the island of Alderney (Fig. \ref{fig1}). Due to the Venturi effect, the Alderney Race is the site of the strongest current in the English Channel. The strong current generates numerous whitecaps, which cover the surface of the race with a constant white coat. The maximum current velocity is about $5\hskip2pt \mathrm{m.s^{-1}}$ and can reach $7\hskip2pt \mathrm{m.s^{-1}}$ under special conditions {\citep{Furgerot&al2020}}. The current is horizontally and vertically sheared by ocean turbulence and surface wave effects (e.g. \cite{Mercier&al2020},\cite{Lopez_PhilTrans20},\cite{Bennis&al2022}). The tidal current moves northeast and southwest during high and low tide, respectively. The flood current is more intense than the ebb current, resulting in a tidal asymmetry of $3\%$ in calm conditions and up to $13\%$ in storm conditions \cite{Bennis&al2022}. Swell, most of the time coming from the Atlantic Ocean, and wind waves propagate over the Alderney Race {\citep{Thiebaut&al2020}}. They are refracted, shoaled and dissipated by the current {\citep{Bennis&al2020}}. The seafloor of the Alderney Race is complex with various morphological features (e.g. submarine cliffs, tidal dunes) and uneven with a wide range of roughness types from coarse sands to rocky blocks of $1\hskip2pt \mathrm{m}$ lengthscale {\citep{Furgerot&al2019}}. Strong tidal currents interact with these bottom features and generate highly energetic three-dimensional turbulence structures that are ejected from the bottom to the surface, as shown by {\cite{Furgerot&al2020}} and simulated by {\cite{Bourgoin&al2020,Bennis&al2021}, for example}. Turbulent structures are visible at the surface of a ship in calm conditions and appear to have a length scale of about $15-20$ m.

\subsection{HF radar facilities}\label{sec2b}

In December 2017, the University of Caen-Normandy deployed two phased-array HF radars (WERA: wave radar), originally developed by {\cite{GGurgel&al1999}}, at the Cape of La Hague, located $5 \hskip2pt \mathrm{km}$ from each other in Goury and Jobourg, respectively (Fig. \ref{fig1}). Each system consists of a 16-element linear receive array, an 8-element square transmit array, and radar electronics. These monostatic radars operate alternately at two different frequencies, $13.5 \hskip2pt \mathrm{MHz}$ and $24.5 \hskip2pt \mathrm{MHz}$, thanks to a 4-element transmit array for $24. 5 \hskip2pt \mathrm{MHz}$ and a 4-element transmit array for $13.5 \hskip2pt \mathrm{MHz}$, while the 16-element receive array is the same for both transmit frequencies. Four sets of measurements per hour are recorded alternately, two sets of $17 \hskip2pt \mathrm{min}$ at $13.5 \hskip2pt \mathrm{MHz}$ and two sets of $12 \hskip2pt \mathrm{min}$ at $24. 5 \hskip2pt \mathrm{MHz}$, allowing an estimation of the surface current with a temporal resolution of about $30\hskip2pt\mathrm{min}$. Given the available frequency bandwidth, the range resolution is $1500\hskip2pt \mathrm{m}$ and $750\hskip2pt \mathrm{m}$ at $13.5 \hskip2pt \mathrm{MHz}$ and $24.5 \hskip2pt \mathrm{MHz}$, respectively. In addition, a temporary allocation of a large frequency bandwidth {around $24.5\hskip2pt\mathrm{MHz}$} allowed testing of a high-resolution mode for one week at a range resolution of $200\hskip2pt \mathrm{m}$. The theoretical angular resolution based on the length of the array is about $14^{\circ}$ at $13.5 \hskip2pt \mathrm{MHz}$ and $7^{\circ}$ at $24.5 \hskip2pt \mathrm{MHz}$.The main characteristics of the HF radars are summarized in table \ref{tab1}. 

\begin{table}[!h]
  \caption{\label{tab1} Overview of the radar characteristics (bandwidth frequency, range, angular and time resolutions, measurement duration) according to the transmission frequency ($13.5 \hskip2pt \mathrm{MHz}$ and $24.5 \hskip2pt \mathrm{MHz}$).}
   \begin{tabular}{|*{3}{l|}}
    \hline
      & $\mathbf{13.5} \hskip2pt \mathrm{MHz}$  & $\mathbf{24.5} \hskip2pt \mathrm{MHz}$  \\ \hline
     \textbf{Bandwidth} ($\mathrm{kHz}$) & 100 & 200 \\ \hline
     \textbf{Range resolution} ($\mathrm{m}$) & 1500 & 750 or 200 \\ \hline
     \textbf{Angular resolution} ($\mathrm{^{\circ}}$) & 14 & 7 \\ \hline
     \textbf{Time resolution} ($\mathrm{min}$) & 30 & 30 \\ \hline
     \textbf{Duration} ($\mathrm{min}$) & 17 & 12 \\\hline
      \textbf{Bragg wavelength $\lambda_B$} ($\mathrm{m}$) & 11.11 & 6.12 \\\hline
      \textbf{Bragg frequency $f_B$} ($\mathrm{Hz}$) & 0.37 & 0.50 \\\hline
       \textbf{Critical current $U_c$} ($\mathrm{m/s}$) & 4.17 & 3.09 \\\hline
     \hline
   \end{tabular}
    \end{table}

\begin{figure}[!h]
\center
\includegraphics[scale=0.5]{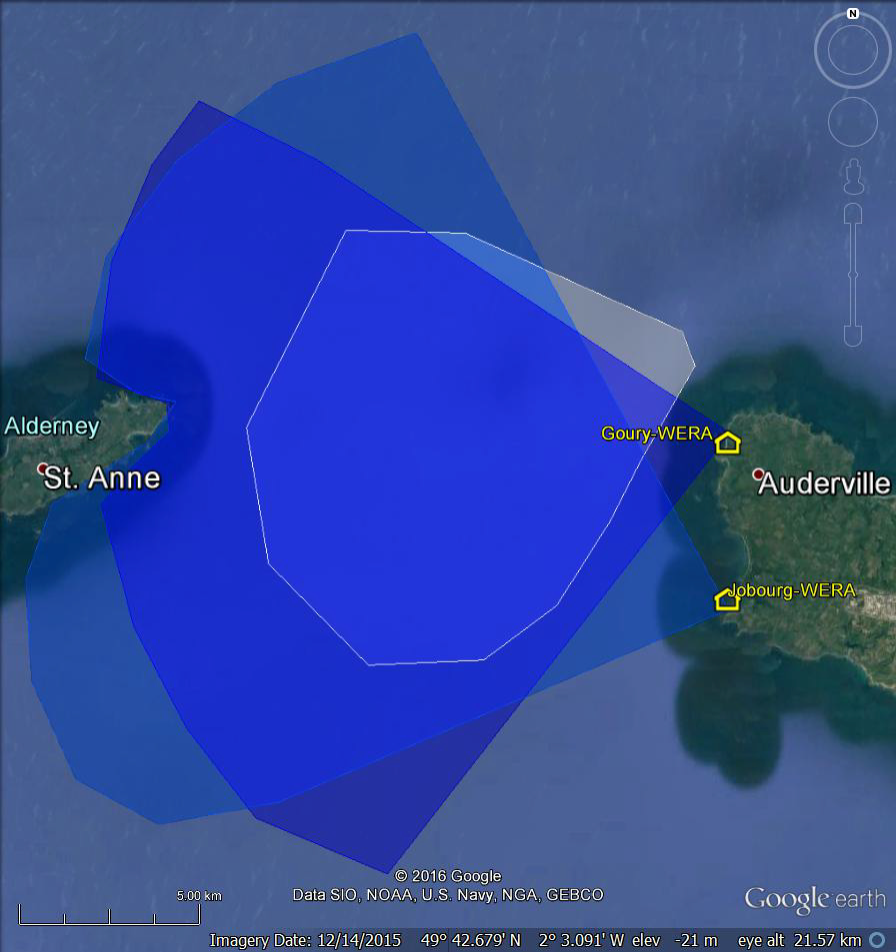}
\caption{\label{fig1}{Locations of the two HF radar sites: Goury-WERA and Jobourg-WERA and their spatial coverage (blue patterns) with the common region (white/gray contours) \citep{helzel_OCEANS17}.}}
\end{figure}

\section{Radar processing}\label{sec3}

The conversion of antenna voltage into surface current maps involves a complex chain of signal processing common to the various types of HF radar. We will briefly recall what is necessary for the following, and refer to e.g. \cite{gurgel_CoastEng99a} for a more detailed description of radar principles. Range gating of the backscattered signal is achieved using standard frequency-modulated continuous wave technology with a chirped transmit waveform. The different range cells of the radar signal are obtained by a simple Fast Fourier Transform (FFT) over the fast time within each transmission cycle.  The theoretical range resolution is related to the single frequency band $B$ by the simple relationship $\Delta R=c/2B$, where $c$ is the speed of light. This allows in principle a resolution of 750 m in the low-resolution mode ($B=200$ kHz) or a resolution of 200 m in the high-resolution mode ($B=750$ kHz). However, the application of a tapered window to prevent secondary lobes in the fast time FFT (which would introduce mutual coupling between the different ranges) reduces the actual range resolution, which is about 2 times coarser.
Complex omnidirectional Doppler spectra for each antenna are further obtained with an FFT on slow time (i.e., at the chirp rate) for each range cell.
The final step is azimuthal discrimination, which requires combining the complex omnidirectional spectra over the antenna array. The most common azimuthal processing method for linear receive arrays is beamforming (BF), which proceeds by delaying and summing the complex antenna signals and can steer the beam continuously in any direction of a half-space. By applying BF to the complex omnidirectional Doppler spectra, azimuthally resolved Doppler spectra (hereafter referred to as directional Doppler spectra) can be obtained for any radar cell. We recall that the directional Doppler spectra are composed of 2 sharp peaks located at plus and minus the Bragg frequency $f_B=\sqrt{g/(2\pi\lambda_B)}$ predicted by the first-order Bragg theory \cite{barrick_AP72}. In the presence of a radial current {($U_r$)}, these peaks are translated by an additional Doppler shift: $\Delta f=U_r/\lambda_B$. The principle of HFR current estimation is based on the accurate measurement of this extra Doppler shift.
It is well known that angular resolution is limited by the size of the array and degrades at large distances. The alternative to BF is the Direction Finding (DF) approach, which is generally used for compact arrays. In the DF technique, each frequency ray of the First-Order Bragg Region (FOBR) in the omnidirectional Doppler spectrum (corresponding to a given value of the radial surface current) is assigned to an azimuthal direction using the MUSIC algorithm \citep{bienvenu_IEEE83,schmidtAP86}. The main idea behind this algorithm is that the vector obtained from the complex antenna gains on the receive array in the direction of a source belongs to the signal subspace (that is, the complementary of the null subspace associated with the zero eigenvalue) of the covariance matrix of the antenna signals. By steering in all possible directions, such an eigenvector can be found by minimizing its projection on the null subspace or, which is equivalent, by maximizing its inverse projection, which is called the MUSIC factor. There is no well-defined criterion for the angular resolution of the MUSIC method, which can be arbitrarily fine depending on the signal-to-noise ratio.
Although DF was originally designed for compact arrays, it has proven to be very valuable for phased-arrays as well. \textcolor{black}{Recently, \cite{dumas_JTECH23} developed an ensemble of techniques to improve both the accuracy and coverage of surface current maps obtained from DF processing of arrays. The results were validated with drifter measurements and shown to outperform BF for 8- and 12-antenna receiver arrays. The main novelty of these techniques is the use of so-called ``antenna grouping'', which consists in applying a DF search algorithm to a large number of sub-arrays, thereby increasing the radar coverage. In addition, an automatic and real-time antenna calibration procedure has been developed in the bistatic configuration, where the remote transmitter is used as an external source to adjust the antenna gains. This last possibility is not available in the monostatic case relevant to this study. Nevertheless, the antenna grouping technique allows to perform a single, manual calibration by systematically searching for the phase corrections that minimize the map rotations between different 3-antenna groups.}

\begin{figure}[!h]
\center
\includegraphics[scale=0.35]{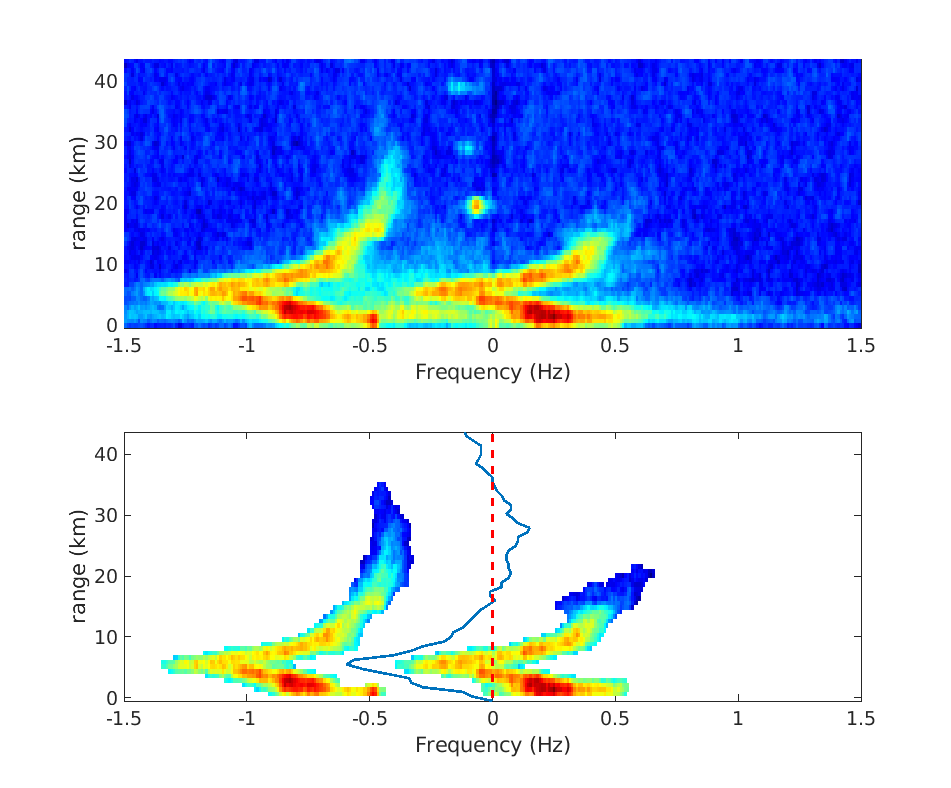}
\caption{Example of a Range-Doppler map obtained with the Jobourg station at 24.5 MHz (750 m resolution mode) in the presence of very strong surface currents. The right Bragg line overlaps the region of negative Doppler shift and the zero-Doppler line (red dashed line) is no longer relevant to identify the ``positive side'' or ``negative side'' of the Bragg regions. A more appropriate separation criterion is obtained with the solid blue line, which is the median of the 2 lines.}
\label{position_points}
\end{figure}

\subsection{Direction Finding for very strong currents\label{secBFDF}}
\textcolor{black}{A key step for the successful application of the DF technique is the correct identification of the FOBR, which is the only relevant frequency interval for the evaluation of the radial surface currents. The classical technique is based on a ``null-finding'' criterion, which amounts to searching for the local minima at both ends of the positive and negative main peaks (see e.g. \cite{rodriguez2022determining} for a recent review), a procedure that requires some parameter tuning and is subject to error. Improved techniques have been developed based on image processing methods \citep{kirincich_JAOT17} that reduce the number of parameters to calibrate and increase the success rate of detection. Another family of techniques proceeds with a statistical approach by making a preliminary estimate of the radial current distribution in a coarsely defined FOBR (with fixed bounds imposed by the maximum possible current) and setting a first-order bound as a quantile of the distribution \citep{dumas_JTECH23}. The bounds are then dynamically adjusted at each observation cycle.  However, the presence of very strong currents raises several issues in the DF processing of Doppler spectra that even these improved techniques cannot handle.  In the case of an extremely large current, the shifted Bragg frequency peak may ``cross'' the zero Doppler line and thus undergo a sign change that makes its identification with the reference Bragg peaks ambiguous. This is illustrated in Figure \ref{position_points}, which shows an azimuthally resolved (i.e. ``beam-formed'') range-Doppler map obtained with BF at the Jobourg station. As can be seen, the extreme variations of the currents in the first 20 km lead to a very unusual shape of the Bragg lines; the right Bragg line, which usually corresponds to a positive Doppler shift, clearly overlaps the region of negative frequencies.  The corresponding critical value for the radial surface current is $U_r=\lambda_B f_B$, which is $4.17$ m/s at $13.5$ MHz and $3.09$ m/s at $24.5$ MHz. These values are generally never reached, except in some very specific locations such as the Alderney Race, where the tidal current can reach such extremely large magnitudes. In the omnidirectional Doppler spectrum, where the contributions from all azimuthal directions are mixed, the occurrence of such phenomena can lead to errors in the identification of the Bragg region, as some Bragg lines may be attributed to the ``wrong'' side of the spectrum. A second problem associated with strong currents is the overlap of the first- and second-order regions for the omnidirectional Doppler spectrum, due to the extreme variation of the first-order Bragg peaks with direction. {These two problems are illustrated in figures \ref{coupe_doppler1} and \ref{coupe_doppler2}, which compare the omnidirectional Doppler spectrum of the Jobourg station with the azimuthally resolved Doppler spectra using BF in two different directions on two different days. It appears that the second order peaks of the latter fall within the main broad peak of the former (Figure \ref{coupe_doppler1}) and that the ``positive'' first order directional Doppler peak lies on the negative side of the frequencies (Figure \ref{coupe_doppler2}).}  Therefore, the identification of the FOBR is problematic and flawed, with errors resulting from misinterpreted second-order contributions. Note that this last problem is specific to DF processing and does not affect BF processing. Therefore, in the case of a phased-array radar, we can take advantage of the BF processing to improve the selection of the FOBR.}
\begin{figure*}[ht]
 \includegraphics[width=1\textwidth]{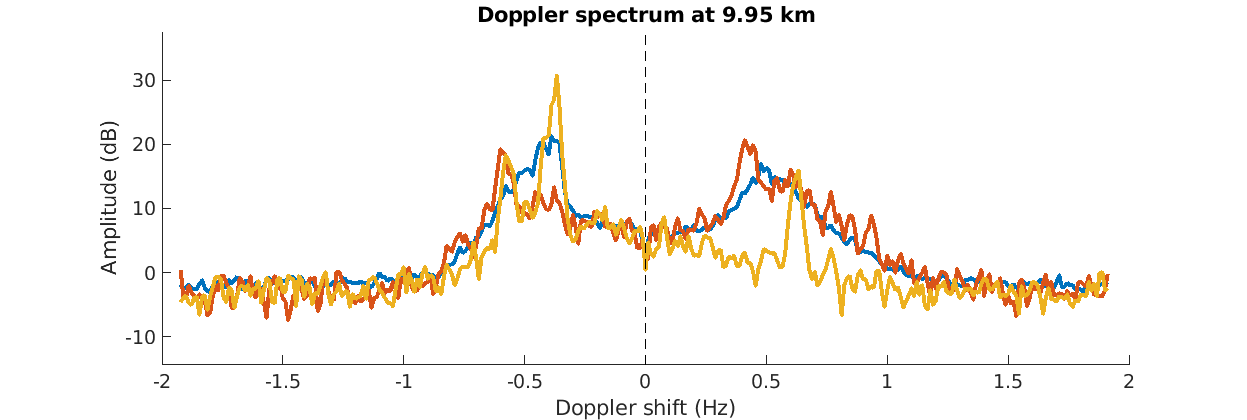}
\caption{Doppler spectra obtained on October 6, 17:20 UTC, 2018 at the Jobourg station (24.5 MHz, 750 m resolution mode) for a range of 10 km. The omnidirectional spectrum (blue line) is compared with the azimuthally resolved spectra at 40$^\circ\mathrm{N}$ CCW (red line) and 100$^\circ\mathrm{N}$ CCW (yellow line). As seen, the second-order peaks of the directional spectra fall inside the main broad peaks of the omnidirectional spectrum, so these second-order contributions can be mistaken with first-order Bragg lines in the DF process.\label{coupe_doppler1}}
\end{figure*}
\begin{figure*}[ht]
  \includegraphics[width=1\textwidth]{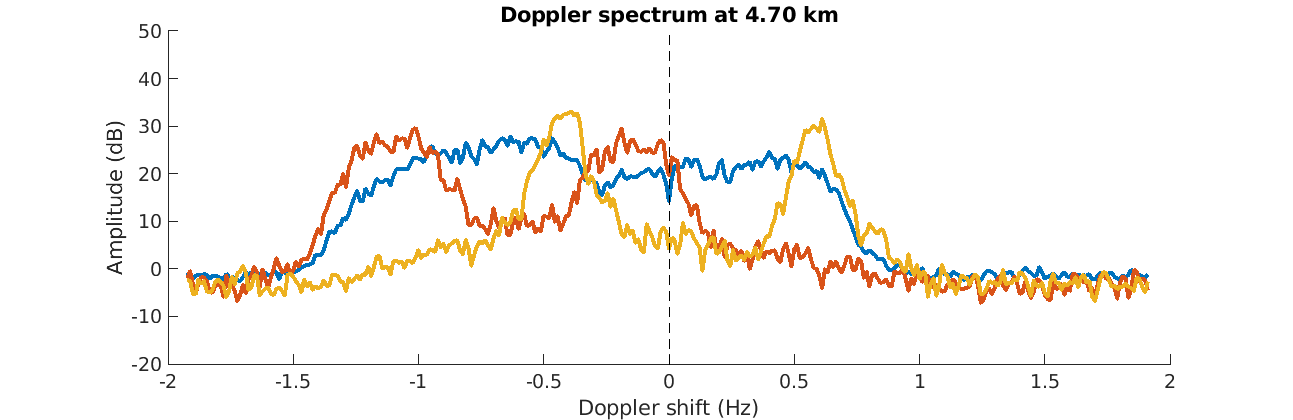}
\caption{Same as Figure \ref{coupe_doppler1} on October 10, 2018, 07:20 UTC at a range of 4.7 km. The radial current in direction 34$^\circ\mathrm{N}$ CCW is so strong that the ``positive'' Bragg peak crosses the zero frequency, making the identification of the Bragg frequencies ambiguous.}
\label{coupe_doppler2}
\end{figure*}
\textcolor{black}{The main idea is to perform a preliminary identification of both sides of the FOBR using directional Doppler spectra. For each range and steering direction, the FOBR can be unambiguously identified within a frequency interval whose width depends on the frequency resolution (hence the integration time), the azimuthal resolution (hence the length of the receive array), and the variation of the radial surface current in the local angular sector.  By sweeping across all steering directions and concatenating the FOBR of each directional Doppler spectrum, we can reconstruct the FOBR of the omnidirectional spectrum without second-order artifacts. The method is illustrated in Figure \ref{detecontour}, which shows the omnidirectional range Doppler map (top) in a situation where both the large currents and the important second-order contribution make it difficult to identify the FOBR with a simple contour detection method. The middle panel shows the selected FOBR, while the bottom panel shows the same using the concatenation of FOBRs obtained with directional Doppler spectra. Although the FOBR selected with a contour criterion is visually intuitive, it is flawed with extra Doppler rays around 10 km (due to misinterpreted second-order or RFI contributions) and misses the strong negative shifts around 5 km due to very strong horizontal current shear.}

\begin{figure}[!h]
\center
\includegraphics[scale=0.32]{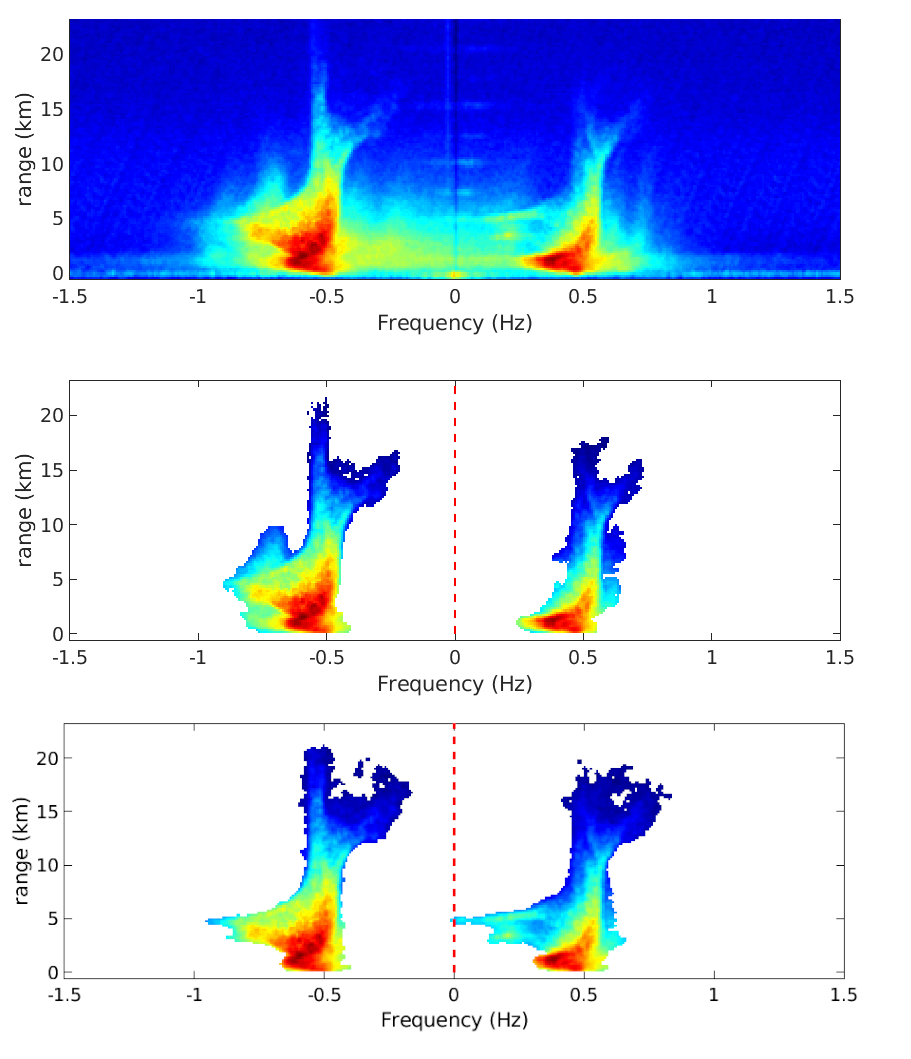}
\caption{Example of range-Doppler map (top panel) for the omnidirectional Doppler spectrum recorded at Jobourg (24.5 MHz, 750 m resolution) in a day with strong waves and currents. The middle panel shows the selected FOBR with a classical contour detection of the omnidirectional spectrum. The bottom panel shows the FOBR obtained by concatenating the FOBR of directional Doppler spectra.}
\label{detecontour}
\end{figure}
\textcolor{black}{While concatenating the FOBR of directional Doppler spectra, we create a table of correspondence between a given Bragg line and the associated steering direction and side of the spectrum (left or right). This helps to improve the DF processing by allowing \textit{a posteriori} control of the Direction of Arrival (DA). Each DA found with the MUSIC algorithm for a given Bragg line in the omnidirectional spectrum is checked to be consistent with the corresponding steering direction(s) found with the BF. This allows the elimination of parasitic DA corresponding to Radio Frequency Interference (RFI), ships, or second-order contributions. From now on, we will refer to the DF processing that uses a preliminary BF processing for the selection of the FOBR and \textit{a posteriori} verification of the DA as the ``Hybrid BF/DF'' method (which should not be confused with another hybridization developed by ACTIMAR called HYDDOA (cf. patent: FR 1562550).}).

\begin{figure*}[ht]
  
\begin{minipage}[c]{0.45\textwidth}
a) \includegraphics[width=1\textwidth]{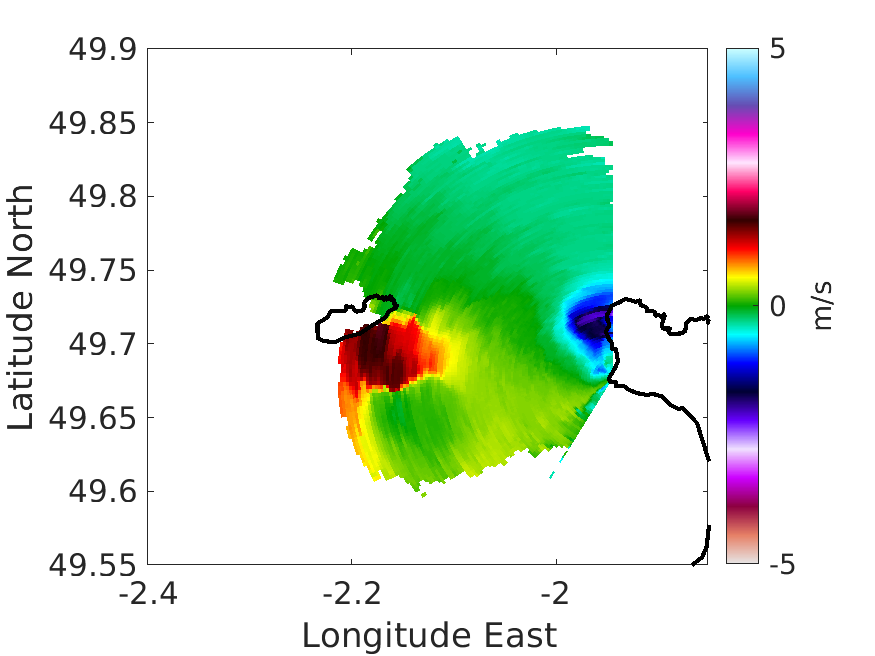}
\end{minipage}
\begin{minipage}[c]{0.45\textwidth}
b)\includegraphics[width=1\textwidth]{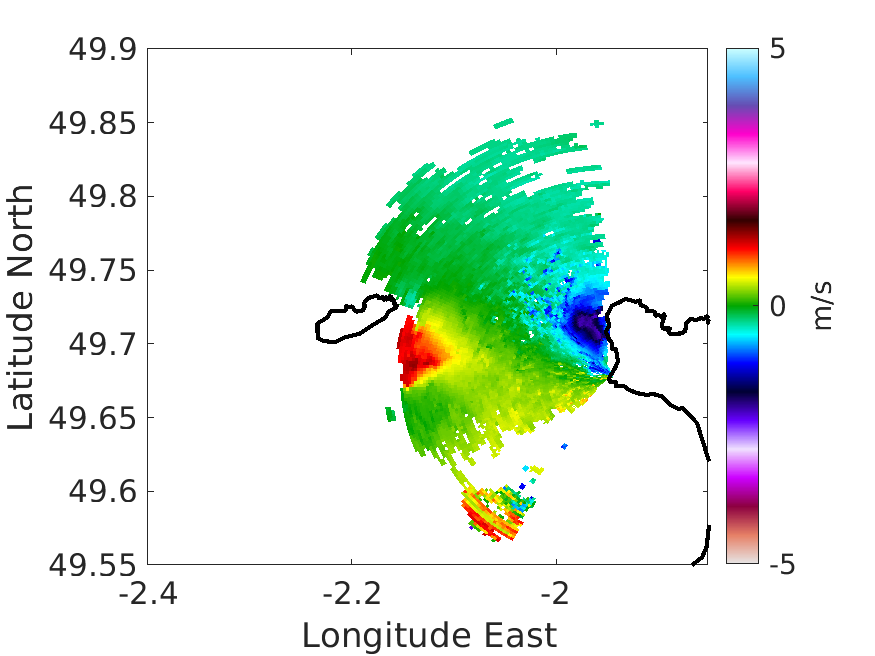}
\end{minipage}

\begin{minipage}[c]{0.45\textwidth}
c)  \includegraphics[width=1\textwidth]{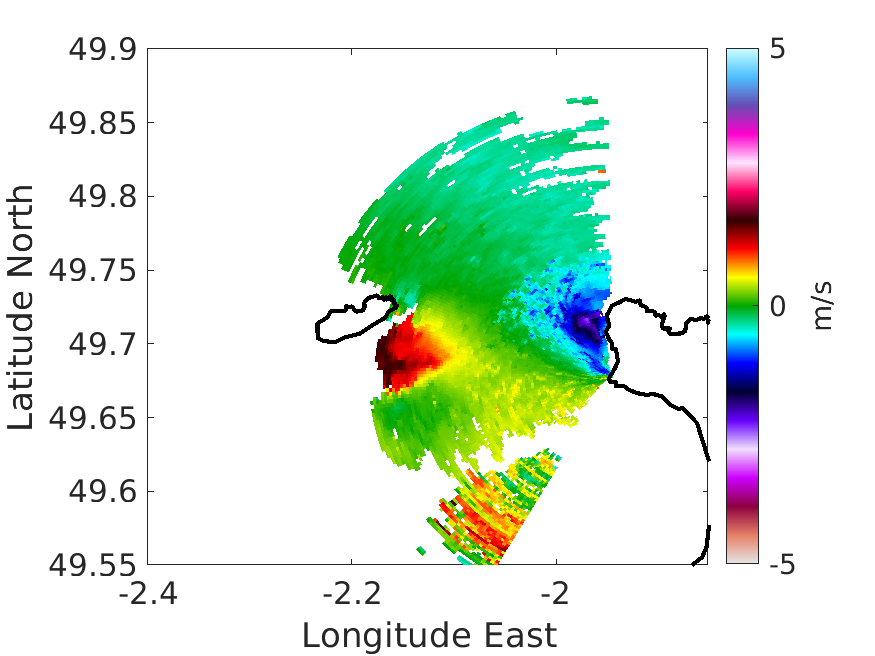}
\end{minipage}
\begin{minipage}[c]{0.45\textwidth}
d) \includegraphics[width=1\textwidth]{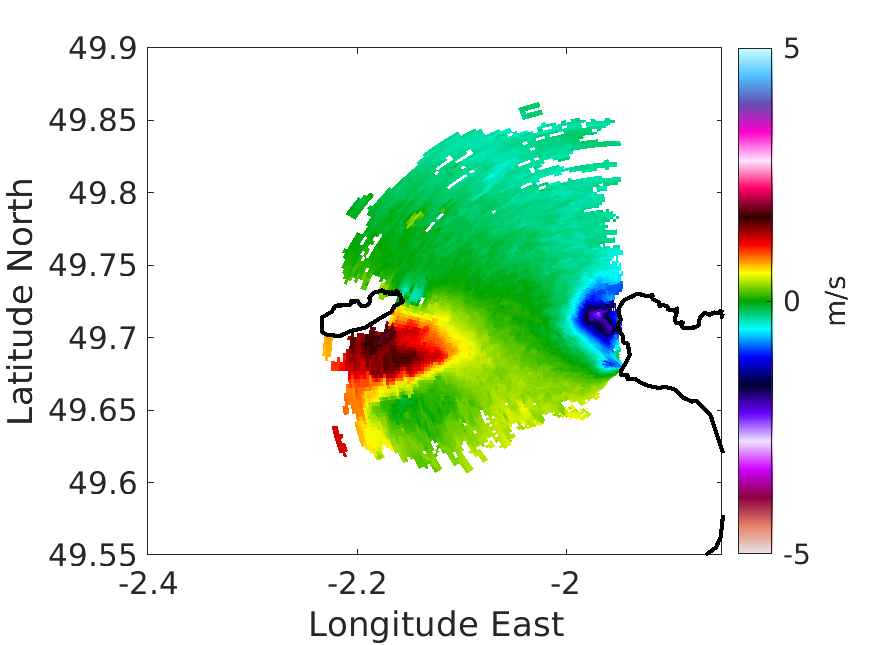}
\end{minipage}

\caption{Radial current maps obtained at the Jobourg station (24.5 MHz, 200 m resolution) on November 14, 03:00, 2020, with different azimuthal processing with the 16-antenna receive array: a) BF b) DF with FOBR selection obtained with dynamical intervals; c) DF with FOBR obtained by contour selection; d) DF with FOBR selection obtained with BF ("hybrid BF/DF").}
\label{match_methodes}
\end{figure*}

\begin{figure*}[ht]
  
\begin{minipage}[c]{0.45\textwidth}
a)\includegraphics[width=1\textwidth]{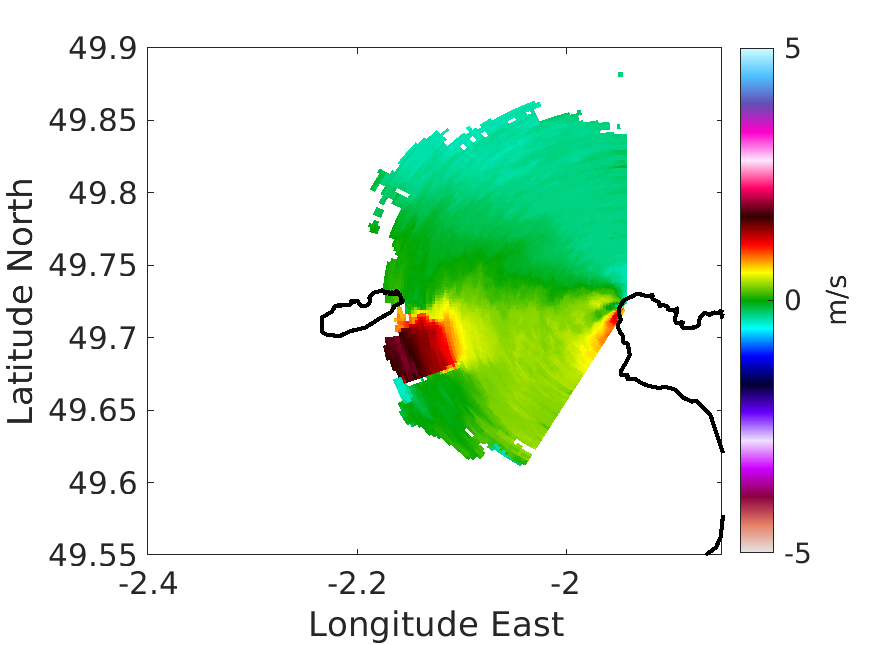}
\end{minipage}
\begin{minipage}[c]{0.45\textwidth}
\hspace{-1.1cm}b)\includegraphics[width=1\textwidth]{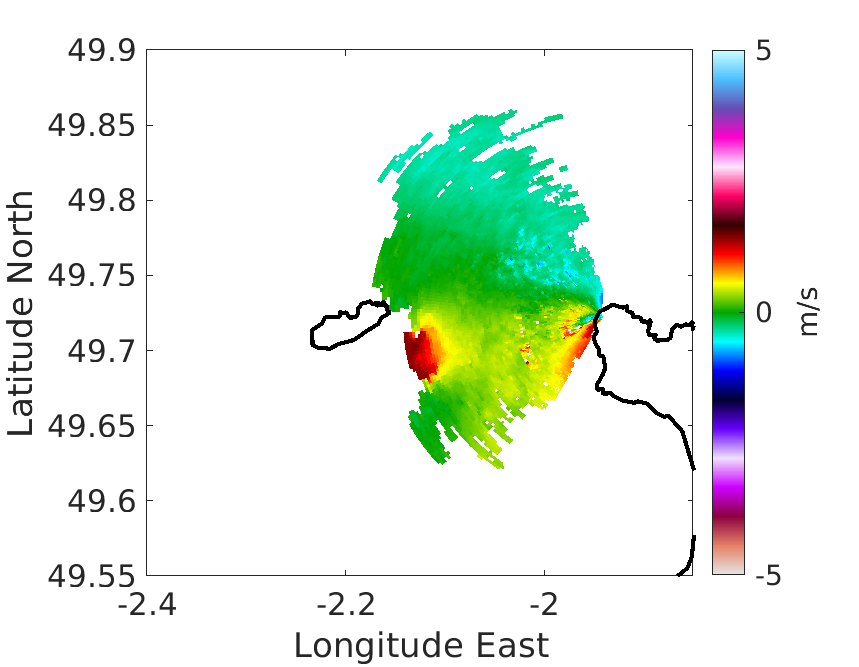}
\end{minipage}
\begin{minipage}[c]{0.45\textwidth}
c)\includegraphics[width=1\textwidth]{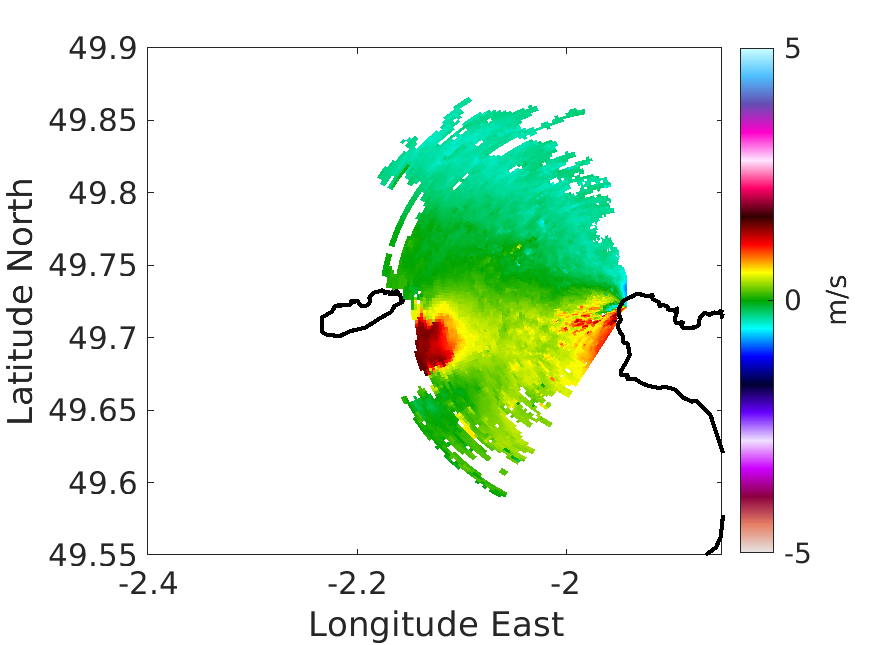}
\end{minipage}
\begin{minipage}[c]{0.45\textwidth}
d)\includegraphics[width=1\textwidth]{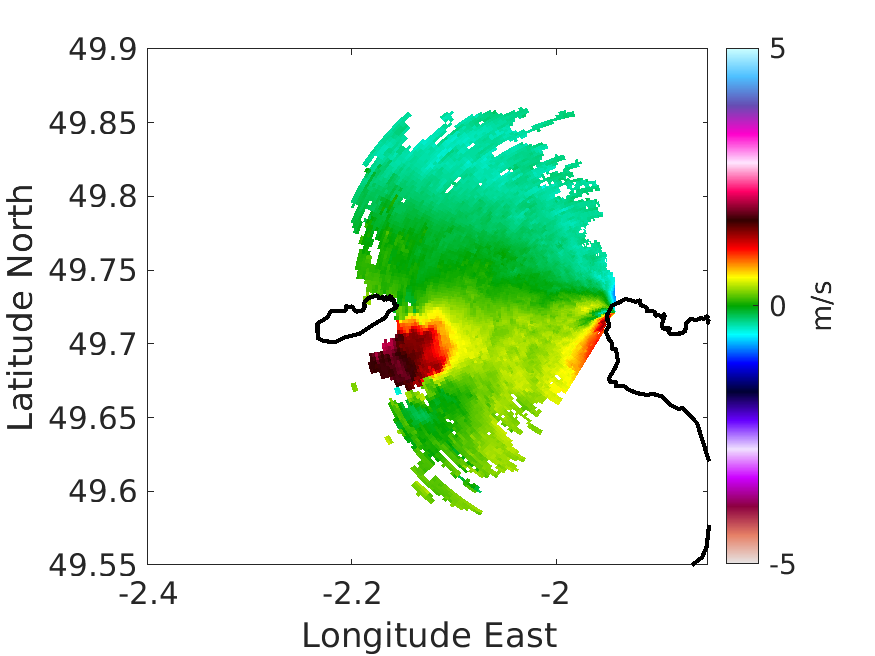}
\end{minipage}

\caption{Same as Figure \ref{match_methodes} for the Goury station.}
\label{match_methodes2}
\end{figure*}

{Figures \ref{match_methodes} and \ref{match_methodes2} show an example of hourly radial maps obtained at the Jobourg and Goury stations using either the classical BF processing or the DF processing with the 3 FOBR selection methods mentioned above (dynamic bounds, contour selection by null detection, or concatenation of the domains obtained with BF). The antenna grouping and noise reduction methods developed in \cite{dumas_JTECH23} were applied to the DF processing. As expected, BF produces a smoother and cleaner map but suffers from angular smearing in the far field and at the edge of the radar coverage where the azimuthal resolution deteriorates. The DF maps are better resolved and show finer spatial structures. However, DF based on dynamic FOBR search interval or contour selection is plagued by isolated spots in the central region of the map as well as spurious structures in the southern part of the map. These artifacts are the remnants of second-order Doppler peaks that have resisted the application of noise removal techniques developed by  \cite{dumas_JTECH23}, due to the overlap between the first and second-order regions in the omnidirectional Doppler spectrum. These artifacts are completely removed by the hybrid BF/DF method. Additionally, this method extends the range and coverage, especially around Alderney Island

Note that the radar coverage obtained with BF is greater at the Jobourg station, while it is similar to that of the Goury station when DF processing is used. This is due to a higher transmitted power at the Jobourg station, which is also affected by a higher level of electromagnetic interference. This is caused by a stronger coupling between the transmitting and receiving antennas (which are closer together) and, in the case of high wind speeds, by gust-induced mechanical vibrations in the antenna arrays, which are fixed to the ground by textile fiber ropes.

\section{Results and performances}\label{sec4}

\subsection{{Performance assessment using a 3D numerical model}}
The validation of HFR measurements is known to be a difficult task, as it is usually based on {\textit{in situ}} measurements that are spatially sparse and have different integration depths and horizontal resolutions. In the absence of a dedicated {\textit{in situ} field campaign}, a surrogate means of validation is comparison with numerical models. Even if they cannot be considered as ground truth, they at least have the advantage of providing spatial maps that allow an overall qualitative comparison (as opposed to a pointwise quantitative comparison with {\textit{in situ} instruments). We have used the {hydrodynamic model, CROCO \citep{Jullien&al2002}, coupled two ways to the spectral wave model, WAVEWATCH-III \citep{WAVEWATCHIII2019}. This coupled model can compute wave-current interactions in three dimensions using the vortex force formalism of \cite{McWilliams&al.2004}. Wave effects on the current are simulated as well as changes in the wave field due to surface currents and sea level. Both models use the same horizontal grids, for which a nesting technique allows the use of three different horizontal resolutions: $600\hskip2pt\mathrm{m}$, $120\hskip2pt\mathrm{m}$, and $30\hskip2pt\mathrm{m}$. CROCO, which runs in three-dimensional mode, also has a vertical discretization based on a following-terrain coordinate $\sigma$ that ranges from $-1$ at the bottom to $0$ at the surface and extends over $20$ levels. The coupled model is forced at its open boundaries by the CST France Atlas \citep{LeroySimon2003} with 114 tidal constituents for tidal dynamics and by two-dimensional wave spectra from the HOMERE numerical wave database for sea state dynamics \citep[e.g.][]{Boudiere&al2013}. At the surface and for both models, wind stress is applied at a spatial resolution of $24.5\hskip2pt\mathrm{km}$ from the ERA5 database \citep{ERA5}. The exchange between CROCO and WAVEWATCH-III is managed by the automatic coupler OASIS \citep{Valcke&al2015}. The numerical configuration used here is based on that of \cite{Bennis&al2020,Bennis&al2022} and \cite{Lopez_PhilTrans20}; which have already been tested on $\textit{in situ}$ and HF radar data. }
The coupled model was run from November 13-19, 2020, a period that covers one week of HFR measurements in high spatial resolution mode (200 m range resolution) and also corresponds to a period of spring tides and strong wind waves during which high surface currents were observed.

The model surface currents obtained in the first child grid at a horizontal resolution of $120$ meters were projected onto the Goury and Jobourg radials for ease of comparison. For a more quantitative evaluation, we decided to focus on the hourly radial current time series observed by HFR in the high-resolution mode (24.5 MHz and 200 m range resolution) at a few strategic locations representative of the site specificity. The selected points, labeled from P1 to P15, are shown in Figure \ref{P1P15} together with the bathymetry of the site. They are representative of the different situations that can occur in the radar domain for the surface current: mid-range positions with a weak Jobourg radial (P01-P03, P08, P10-P11), near-range positions with a strong Jobourg radial (P04-P07), far-range in deeper water (P09, P12), and far-range around the island of Alderney. The HFR-based hourly currents were calculated using both the BF and hybrid BF/DF processing. To evaluate the two HFR radials separately, the surface current vectors calculated with the CROCO model were projected along the two radial directions as seen from the Goury and Jobourg stations. To quantify the amplitude of variation of the radials, we define a suitable statistical index $A$ as the mean difference (during the time considered) between the maximum and minimum value of the current {during all tidal cycles}:
\be
  A=\langle (U_r)_{max}- (U_r)_{min} \rangle
  \ee
 The discrepancy between the model and HFR based radials has been quantified, on the one hand, by their RMS difference {($\sigma_{BF}$ and $\sigma_{DF}$)} and their Pearson correlation coefficient {($\rho_{BF}$ and $\rho_{DF}$)} and, on the other hand, by their difference in amplitude $A$. These statistical indicators are given in tables \ref{tab2} and \ref{tab3} for the radial velocities measured at the Jobourg and Goury stations. This table shows that the agreement between the model and the radar measurements, or between the two types of radar processing (BF or DF), is generally very good, except for a few locations. {For the Jobourg station and using the BF method, $\sigma_{BF}$ varies between $0.13\hskip2pt\mathrm{m.s^{-1}}$ and $0.74\hskip2pt\mathrm{m.s^{-1}}$ with a mean $\sigma_{BF}$ around $0.37\hskip2pt\mathrm{m.s^{-1}}$. Close values are found using the DF method with a $\sigma_{DF}$ between $0.16\hskip2pt\mathrm{m.s^{-1}}$ and $0.76\hskip2pt\mathrm{m.s^{-1}}$ and a mean $\sigma_{DF}$ around $0.35\hskip2pt\mathrm{m.s^{-1}}$. Maximum values for $\sigma_{DF}$ and $\sigma_{BF}$ are observed in P14, which is at the limit of spatial coverage as shown in Fig. \ref{velocityMap.HFRvsModel}. These values remain weak relative to the magnitude of the flow for this period. The same behavior is observed for the Goury station, with a $\sigma_{DF}$ in [$0.17\hskip2pt\mathrm{m.s^{-1}}$ ; $0. 88\hskip2pt\mathrm{m.s^{-1}}$] and a $\sigma_{BF}$ in [$0.17\hskip2pt\mathrm{m.s^{-1}}$ ; $0.86\hskip2pt\mathrm{m.s^{-1}}$]. The mean values of $\sigma_{BF}$ and $\sigma_{DF}$ are around $0.38\hskip2pt\mathrm{m.s^{-1}}$ and $0.4\hskip2pt\mathrm{m.s^{-1}}$, respectively. As for the Jobourg station, the maximum values of $\sigma$ were found at point P13, located near the island of Alderney at the limit of the HFR spatial coverage. The overall agreement between model and radar measurements is good at Jobourg with 8 out of 15 values of $\rho_{BF}$ and $\rho_{DF}$ greater than or equal to $0.95$ and very good at Goury with 11 out of 15 values greater than or equal to $0.95$. These values are consistent with the previous analysis by \cite{Lopez_PhilTrans20} using the standard BF method and the fully coupled 3D model, MARS-WAVEWATCH-III, by \citep{Bennis&al2020,Bennis&al2022}. In this study, the RMSE was calculated by spatio-temporal averaging over the radar coverage at different tidal cycles and was found to range from $0.18\hskip2pt\mathrm{m.s^{-1}}$ to $0.60\hskip2pt\mathrm{m.s^{-1}}$. In tables \ref{tab2} and \ref{tab3} we have highlighted in blue the points with the best agreement (P4 and P10, with less than 10\% in the statistical parameters) and in red those with the maximum discrepancy (P1, P5, P7, and P9, with more than 20\% difference in the statistical parameters) for the 2 radar stations simultaneously. Since these last four locations are within the limits of spatial coverage, this larger discrepancy is not due to outliers in the HFR estimate. \cite{Lopez_PhilTrans20} pointed out that the lowest comparison values occur in shallow areas with a strong bathymetric gradient. The impact of these steep gradients is confirmed at locations P1, P5, and P7, in the vicinity of which the induced heterogeneity of radial velocities may not be well captured by the model, which uses a horizontal resolution of 120 meters.
  
    Examples of the corresponding time series are shown in figures \ref{figP4P10}, \ref{fig:P5}, and \ref{fig:P7}, together with the simultaneous value of the significant wave height obtained with the numerical model at the same location. Although the low-resolution wind forcing of the ERA5 reanalysis may lead to an imperfect estimate of the significant wave height, which is also affected by the surface current and sea surface height, it is sufficient to provide the trend and magnitude of the latter. The model and the HFR radial velocity are in better agreement for the Goury radial velocity than for the Jobourg station. This can be attributed to the aforementioned practical limitations of the antenna receiving array at Jobourg (section \ref{secBFDF}), namely a higher level of noise in the radar signal and a greater sensitivity of the antennas to wind gusts, which can induce mechanical vibrations in the antenna arrays. This could explain some of the discrepancies when the significant wave height exceeds $2\hskip2pt\mathrm{m}$. Finally, the discrepancies may be related to the turbulence modeling in the numerical model.


    The model radial velocities show smoother variations than observed with the radar. This may be due to the use of the LES Smagorinsky turbulence model for horizontal mixing and the RANS-GLS turbulence model for vertical mixing, which are known to smooth the flow compared to other types of turbulence models such as the $\alpha$ model \citep{Bennis&al2021}. 
    
    Figure \ref{velocityMap.HFRvsModel} shows two different tidal events on November 15, where the flow velocity is most intense. For these two events, the DF method can produce i) a gap-free map thanks to the \cite{dumas_JTECH23} group antenna method, and ii) a velocity field showing significant horizontal shear, which is traditionally difficult to measure and extract from HFR data. The region of intense velocities is less extensive than in the numerical simulations but follows bathymetric contours as previously reported in the literature. Note the reduction in radar coverage at the ebb. This corresponds to a stronger sea state around this time (as seen from the evolution of the significant wave height), which can significantly degrade the working range of the radar \citep{halverson_JTECH17}.


  
  \begin{table*}[!h]
    \caption{\label{tab2} Amplitude of variation (obtained as the mean difference between the maximum and the minimum over a tidal cycle) of the radial current estimated at the Jobourg station with the CROCO model ($A_m$); the HFR-based BF estimation ($A_{BF}$); the HFR-based DF estimation ($A_{DF}$). RMS difference and Pearson correlation coefficient between the CROCO model and the HFR-based radial currents estimated with BF ($\sigma_{BF}$)and $\rho_{BF}$) and DF ($\sigma_{DF}$ and $\rho_{DF}$), respectively. All values are given in m/s. The model amplitudes that have a discrepancy larger than 20\% (respectively, smaller than 10\%) have been highlighted in red (respectively, blue). The RMS differences that are smaller than 10\% of the model amplitude and the correlation coefficients that are larger than 0.95 have been highlighted in blue. }
   \end{table*}

   \begin{center}
  \footnotesize{  \begin{tabular}{|l|l|l|l||l|l|l|l|}
    \hline
      & $A_m$ & $A_{BF}$ & $A_{DF}$ & $\sigma_{BF}$ & $\sigma_{DF}$ & $\rho_{BF}$ & $\rho_{DF}$\\ 
      \hline    
\textbf{P1} & \rouge 1.81 & 1.11 & {1.29}& 0.34 & 0.33 & 0.95 & 0.92\\ 
\hline
\textbf{P2} & 0.79 & {0.84} & 0.94& {0.13} & 0.16 & 0.89& 0.85\\ 
\hline
\textbf{P3} & 1.03 & 1.04 & 1.02& 0.20 & 0.20& 0.86& 0.86\\ 
\hline
\textbf{P4} &  \bleu 6.17 & 5.52 & 5.49& \bleu 0.50 & \bleu 0.49&  \bleu 0.99&  \bleu 0.99 \\ 
\hline
\textbf{P5} & \rouge 7.28 & 6.71 & 6.68& 0.66 & {0.58} & 0.99& 0.99\\ 
\hline
\textbf{P6} & 6.57 & 6.87 & 6.89& 0.48 & {0.42} & 0.99 & 0.99\\ 
\hline
\textbf{P7} & \rouge 4.54 & 3.72 & {4.10}& 0.46 & {0.38}&0.99&0.99 \\ 
\hline
\textbf{P8} & 2.67 & 2.25 & 2.25& 0.28 & 0.27 &0.99&0.99\\ 
\hline
\textbf{P9} & \rouge 1.73 & 1.26 & 1.37& 0.25 & 0.21 &0.95 & 0.95\\ 
\hline
\textbf{P10} &  \bleu 2.00 & {2.06} & 2.20&  \bleu 0.22 &  \bleu 0.20 & \bleu 0.96&  \bleu 0.97 \\ 
\hline
\textbf{P11} & \rouge 0.77 & 1.13 & {0.96}& 0.18 & 0.20 &0.81& 0.77\\ 
\hline
\textbf{P12} & 1.28 & {1.12} & 1.02& 0.28 & 0.26 &0.65& 0.68 \\ 
\hline
\textbf{P13} & 3.28 & {3.46} & 3.51& 0.47 & 0.48&0.94& 0.94\\ 
\hline
\textbf{P14} & 3.29 & 3.47 & 3.45& 0.74 & 0.76 &0.82& 0.81\\ 
\hline
\textbf{P15} & 3.58 & 3.12 & {3.27}& 0.45 & {0.41}&0.95& 0.95 \\ 
\hline

   \end{tabular}}
    \end{center}
  
\begin{table*}[!h]

  \begin{center}
    \caption{\label{tab3} Same as Table \ref{tab3} for the Goury station.}
  \footnotesize{     \begin{tabular}{|l|l|l|l||l|l|l|l|}
    \hline
  & $A_m$ & $A_{BF}$ & $A_{DF}$ & $\sigma_{BF}$ & $\sigma_{DF}$ & $\rho_{BF}$ & $\rho_{DF}$\\ 
 \hline

\textbf{P1} & \rouge 2.72 & 3.55 & 3.66&  0.28 & 0.34 &0.98& 0.97\\ 
\hline
\textbf{P2} & 5.13 & 4.80 &  4.99& 0.44 &  0.42&0.99& 0.98 \\ 
\hline
\textbf{P3} & 4.77 &  4.12 & 3.95&  0.48 & 0.54 &0.99&0.99\\ 
\hline
\textbf{P4} & \bleu 2.59 & 2.91 & 2.91&  \bleu 0.18 & \bleu 0.25&\bleu 0.99& \bleu 0.98\\ 
\hline
\textbf{P5} & \rouge 2.14 &  2.45 & 2.62&  0.17 & 0.23 &0.98&0.98\\ 
\hline
\textbf{P6} & 2.99 & 2.31 &  2.46& 0.49 &  0.46 &0.99&0.99\\ 
\hline
\textbf{P7} & \rouge 2.02 &  2.30 & 2.48& 0.22 & 0.22&0.97& 0.97\\ 
\hline
\textbf{P8} & 4.68 &  4.71 & 4.09& 0.35 & 0.36&0.99& 0.99\\ 
\hline
\textbf{P9} & \rouge 0.54 & 0.85 &  0.72& 0.19 &  0.17 &0.69&0.65\\ 
\hline
\textbf{P10} &  \bleu 4.19 &  3.78 & 4.03& \bleu 0.32 &  \bleu 0.31& \bleu 0.99& \bleu 0.99 \\ 
\hline
\textbf{P11} & 4.54 & 4.32 &  4.54& 0.37 &  0.35& 0.99&0.98 \\ 
\hline
\textbf{P12} & 2.08 &  2.06 & 1.98& 0.21 & 0.21&0.95& 0.96\\ 
\hline
\textbf{P13} & 4.91 & 4.40 &  4.64& 0.86 & 0.88&0.87& 0.88 \\ 
\hline
\textbf{P14} & \rouge 4.12 & 1.60 &  2.22& 0.72 & 0.84&0.92& 0.91\\ 
\hline
\textbf{P15} & \rouge 4.18 & 1.58 &  1.94& 0.49 &  0.42&0.90& 0.94\\ 
\hline


   \end{tabular}}
    \end{center}

   \end{table*}


\begin{figure*}[!h]
\center
\includegraphics[scale=0.9]{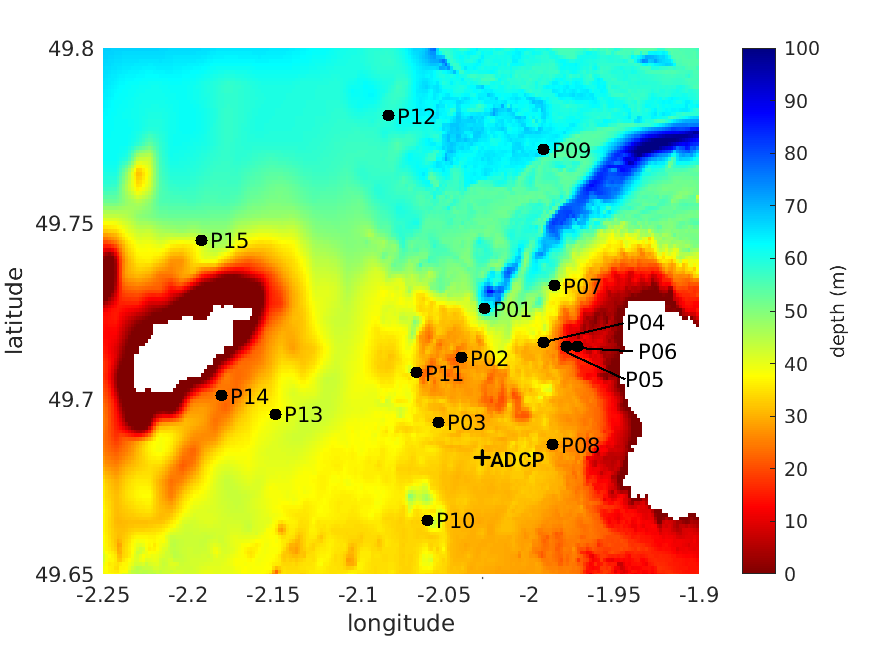}
\caption{Location of the points P1 to P15 for the evaluation of the HFR-based time series of radial currents. The ADCP location is marked with a cross. The {bathymetric depth} is superimposed in color scale.}
\label{P1P15}
\end{figure*}

\begin{figure*}[!h]
\includegraphics[scale=0.25]{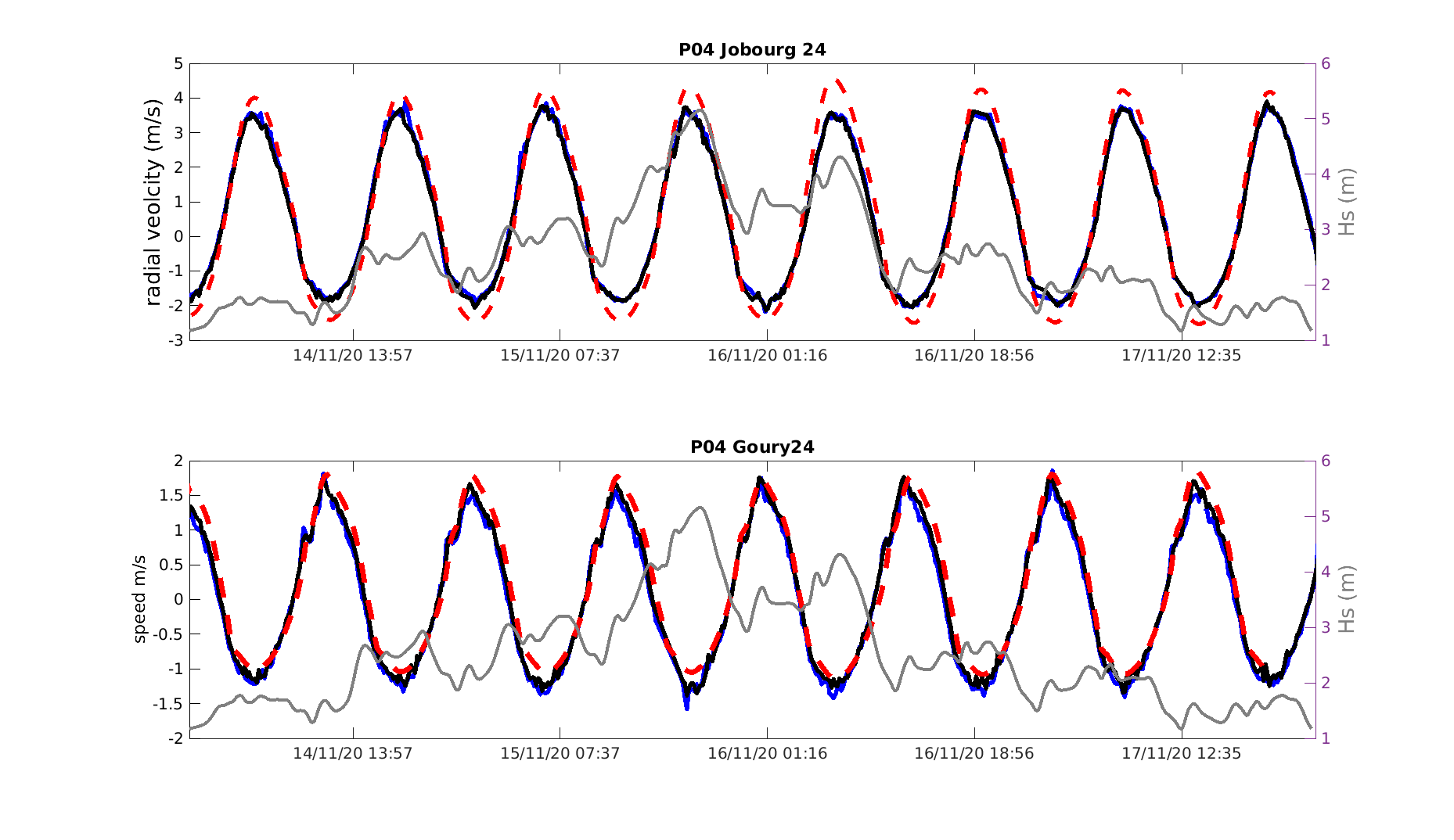}
\caption{High-resolution Radial currents obtained at location P4 from HFR measurements (24.5 MHz, 200 m resolution mode) with DF (blue solid lines) and BF (black solid lines). A comparison is given with the radial current inferred from the CROCO{-WAVEWATCH-III} model (gray solid lines {for the significant wave height and red dashed lines for the radial velocity})\label{figP4P10}}
\end{figure*}

\begin{figure*}[!h]
\includegraphics[scale=0.25]{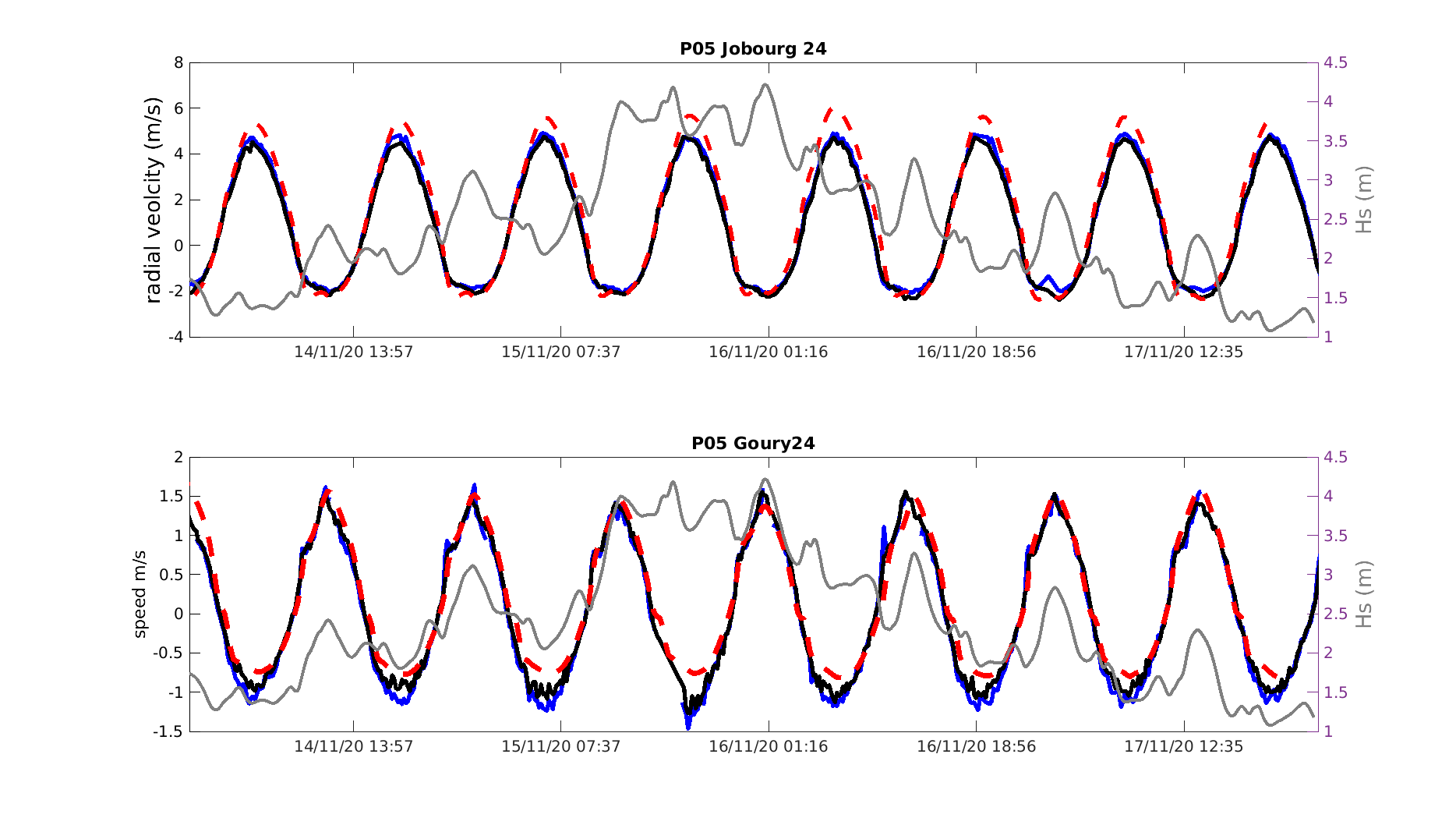}
\caption{Same as Figure \ref{figP4P10} for the radial currents obtained at location P5. }\label{fig:P5}
\end{figure*}

\begin{figure*}[!h]
\includegraphics[scale=0.25]{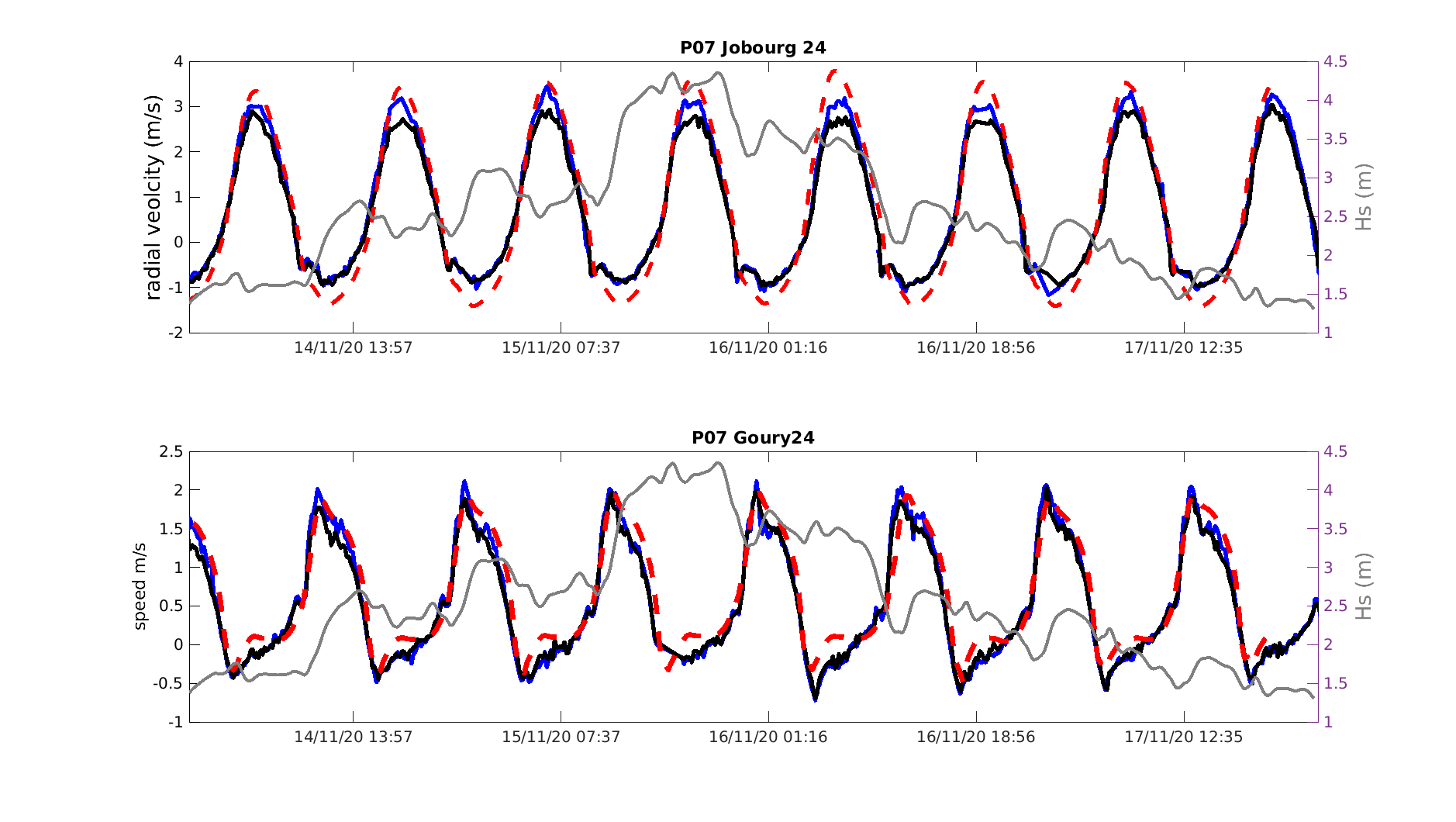}
\caption{Same as Figure \ref{figP4P10} for the radial currents obtained at location P7.}\label{fig:P7}
\end{figure*}

\begin{figure*}[!h]
\begin{minipage}[c]{0.5\textwidth}
\includegraphics[scale=0.32]{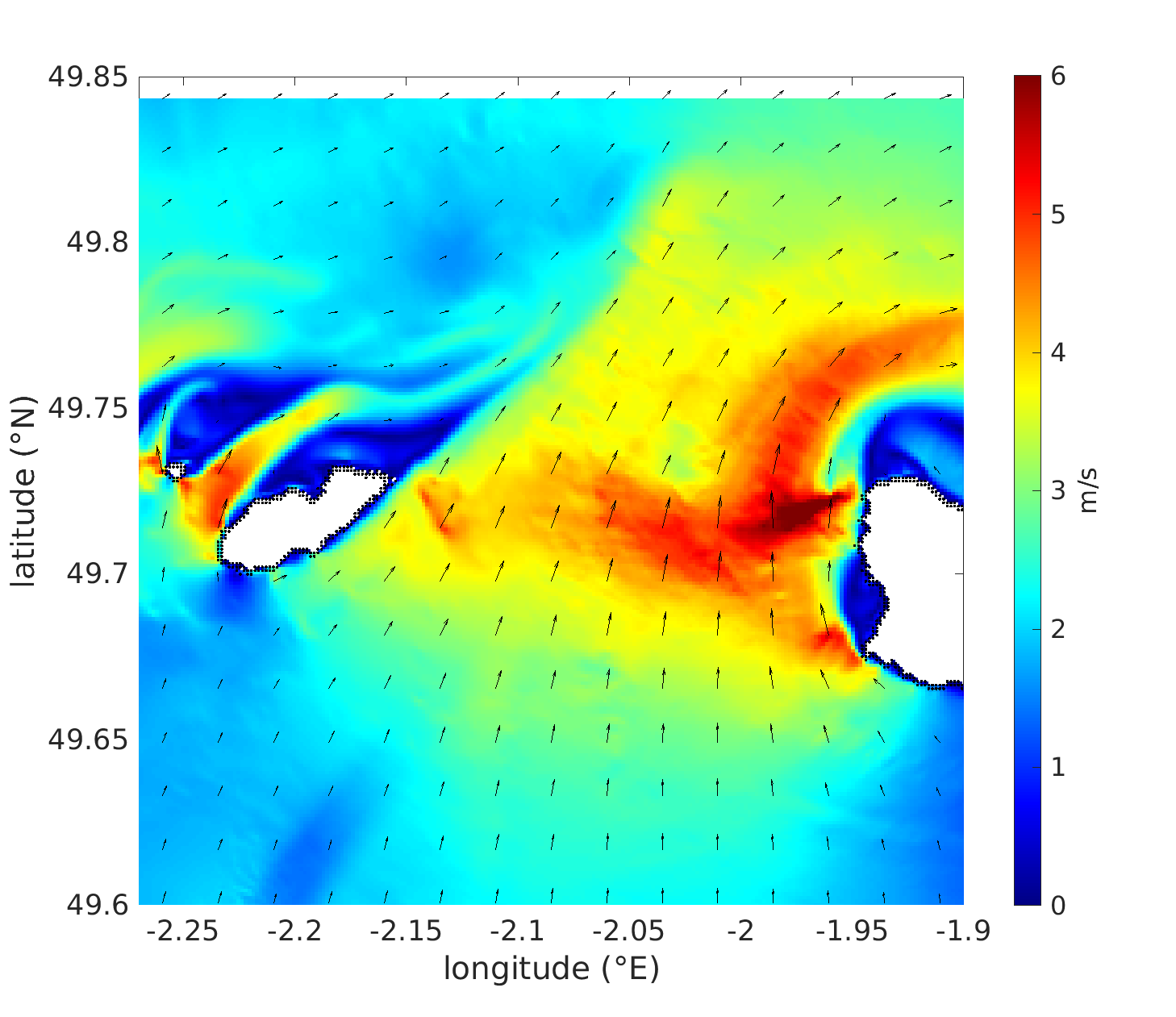}
\end{minipage}\begin{minipage}[c]{0.5\textwidth}
\includegraphics[scale=0.275]{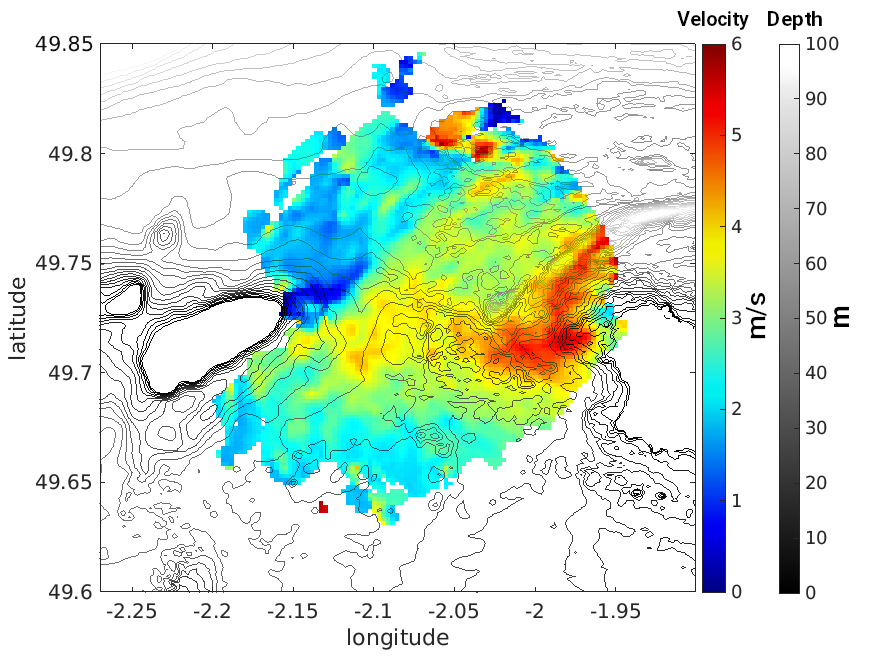}
\end{minipage}
\begin{minipage}[c]{0.5\textwidth}
\includegraphics[scale=0.325]{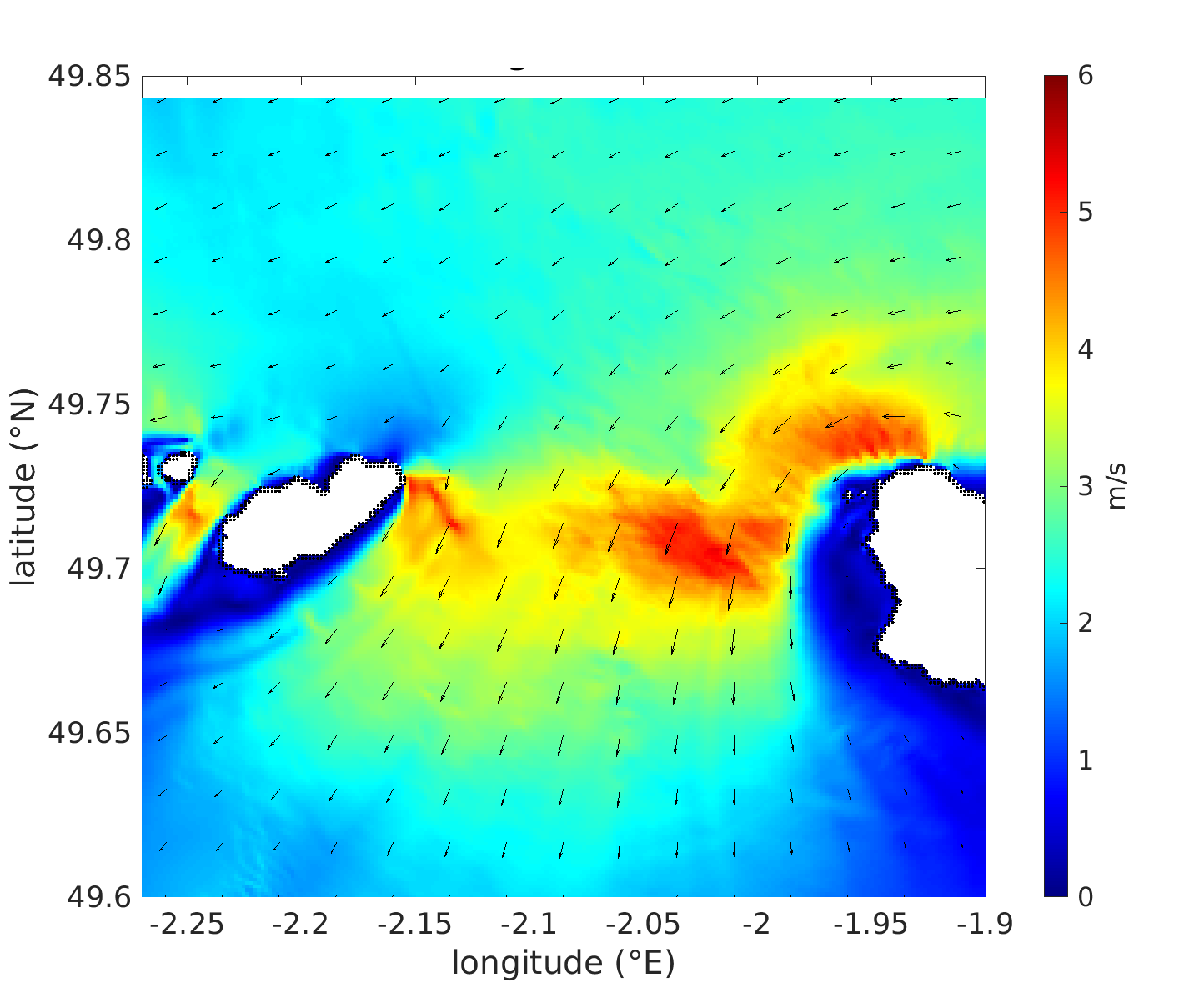}
\end{minipage}\begin{minipage}[c]{0.5\textwidth}
\includegraphics[scale=1.05]{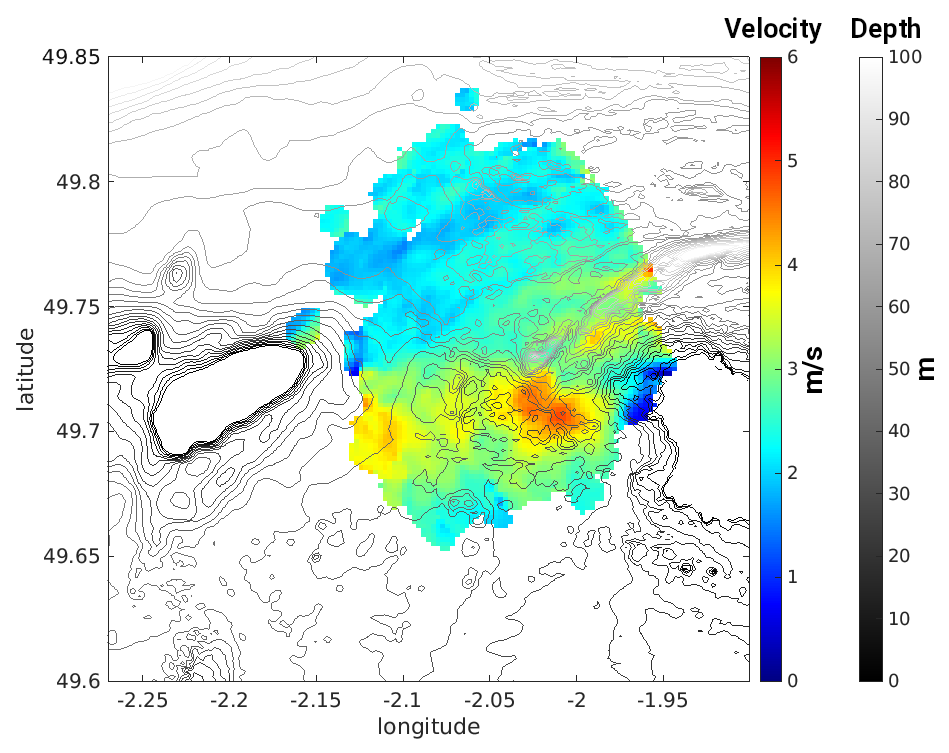}
\end{minipage}
\caption{\label{velocityMap.HFRvsModel}{Magnitude of the total current (color scale) with superimposed bathymetric contours (gray scale) and velocity vectors (black arrows). Results are obtained from the CROCO model (left panel) and the HFR measurements in the high-resolution 200 m mode (right panel) using DF method for flood (first row) and ebb (second row) flow}.}
\end{figure*}

\subsection{{Performance assessment using \textit{in situ} measurements}}

Another way to assess the HFR current is by comparison with {Acoustic Doppler Current Profiler (ADCP)} measurements. A bottom-mounted ADCP was installed in 2017 and maintained for one year to measure the {vertical} profile of the current at a depth of 30 to 40\hskip2pt$\mathrm{m}$ near point P10 at position {-2.0296$^\circ$E} and {49.6808$^\circ$N}. It measures the {three components of flow velocity} from the bottom {to the surface} every meter (which is the bin size) after the blanking zone \citep{Furgerot&al2020}. The measurement cell located 2 meters below the water surface, taking into account the tidal variations, was used to make a relevant comparison with the HFR measurement and to avoid spurious reflections from the sea surface {echo} contaminating the closest bin to the surface. The ADCP measurements are updated every 10 minutes, with an observation time of 20 minutes, which is comparable to the integration time required for the HFR observations (12 minutes). Since ADCP velocities were not available during the HFR period, another period with ADCP in the field and similar tidal conditions was chosen for comparison. Thus, HFR data from November 13 to 18, 2020 were tested against ADCP data from November 3 to 8, 2017. The two periods differ only in the sea state, which was less energetic in November 2017.
The entire ADCP period is covered by the equivalent HFR cycle, except for two days of missing data (Fig. \ref{ADCPvsRadar}). 

A very good agreement with the ADCP data is observed with the two radar signal processing methods (BF and DF), which give similar surface velocities. The RMSE and Pearson correlation coefficients are found to be 0.27 m/s and 0.96 with BF, while they are equal to 0.26 m/s and 0.97 with DF. The largest discrepancy is observed for the maximum values of the absolute velocity, which is known to be the most difficult to estimate with HFR radar processing algorithms, especially for the ebb current \citep{Lopez_PhilTrans20}. These differences are also explained by the different spatial scales involved in the two types of measurements, namely the size of the integration cells for HFR and ADCP data (about a kilometer versus a few meters) and the effective sensing depth (2 m below the sea surface for ADCP, about 30 cm for radar). Finally, the pointwise nature of the ADCP at a specific geographic position may account for some local effects that are not present in the HFR data.  Despite these structural differences, the agreement between the two types of measurements remains remarkably good, with an RMSE of less than 10\% of the typical amplitude.

\begin{figure}[!h]
\center
\includegraphics[scale=0.3]{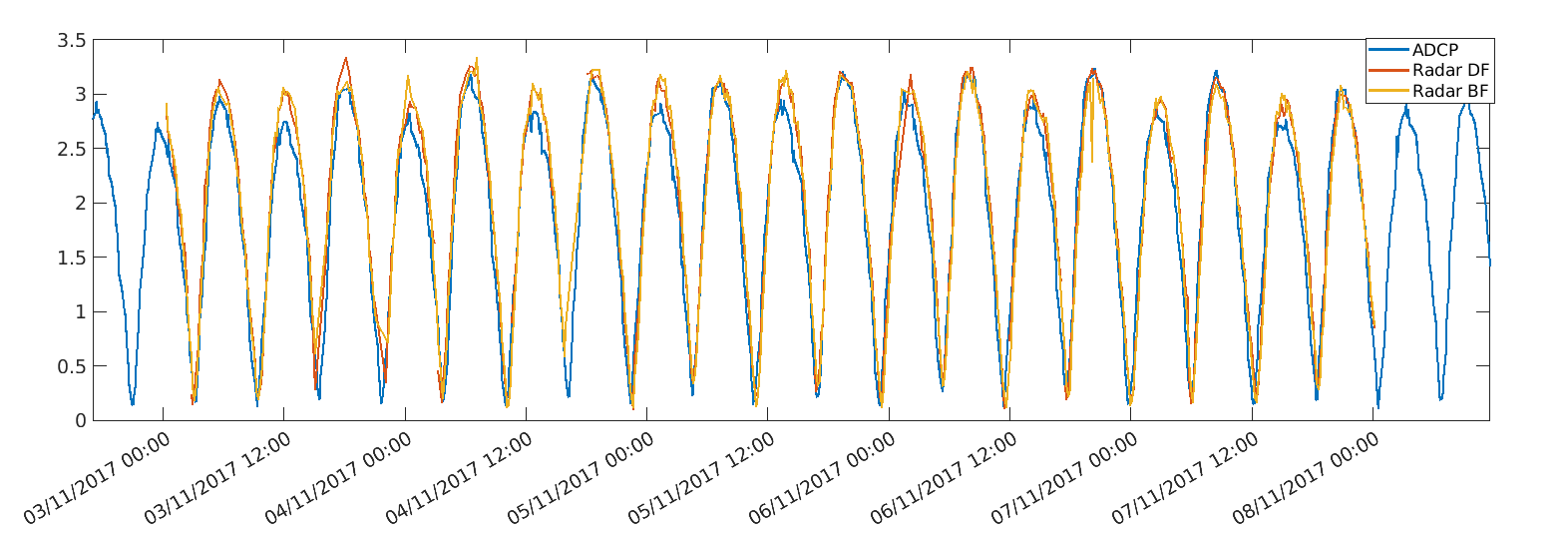}
\caption{\label{ADCPvsRadar}Magnitude of the total current measured by the ADCP at 2 m below the surface {(blue solid line) and} the high-resolution HFR measurement (24.5 MHz, 200 m range resolution) using BF {(yellow solid line) and} using DF {(red solid line)}.}
\end{figure}

\subsection{The influence of range resolution}

As seen above, the strong horizontal shear of the surface flow at some locations requires a high spatial resolution in both range and azimuth to be captured by the HFR measurement. The dual frequency capability (13 MHz and {24.5 MHz}) of the HFR network as well as the dual resolution mode at the 24.5 MHz frequency (750 m or 200 m) provide the opportunity to test this effect. The first observation is the narrowness of the first-order Bragg peaks in the high-resolution radar data. Figure \ref{HFRResol} shows {three} Doppler energy spectra at 200 m range resolution in adjacent cells near point P5. Superimposed is the average of these spectra, which in principle represents the Doppler spectrum that would be recorded in a low-resolution {1500 m cell.} As can be seen from these plots, the low-resolution first-order Bragg peak {(1500 m)} is much broader than its high-resolution counterpart (200 m) and presents several maxima that can be misleading when applying azimuthal processing. DF and BF can therefore select different values for the peak frequency, resulting in different surface velocities. This suggests that the shearing and meandering of the surface current in the Alderney Race has a characteristic scale that is much smaller than the typical radar resolution of 1500 m and can only be captured in the high-resolution mode of 200 m, where the broadening of the Doppler peak is less pronounced, if not absent. {This opens the way to characterize the horizontal shear by assessing the relevant length scale of each meander through its representation in the Doppler spectrum. }

\begin{figure}[!h]
\label{fig:fourchette}
\includegraphics[scale=0.2]{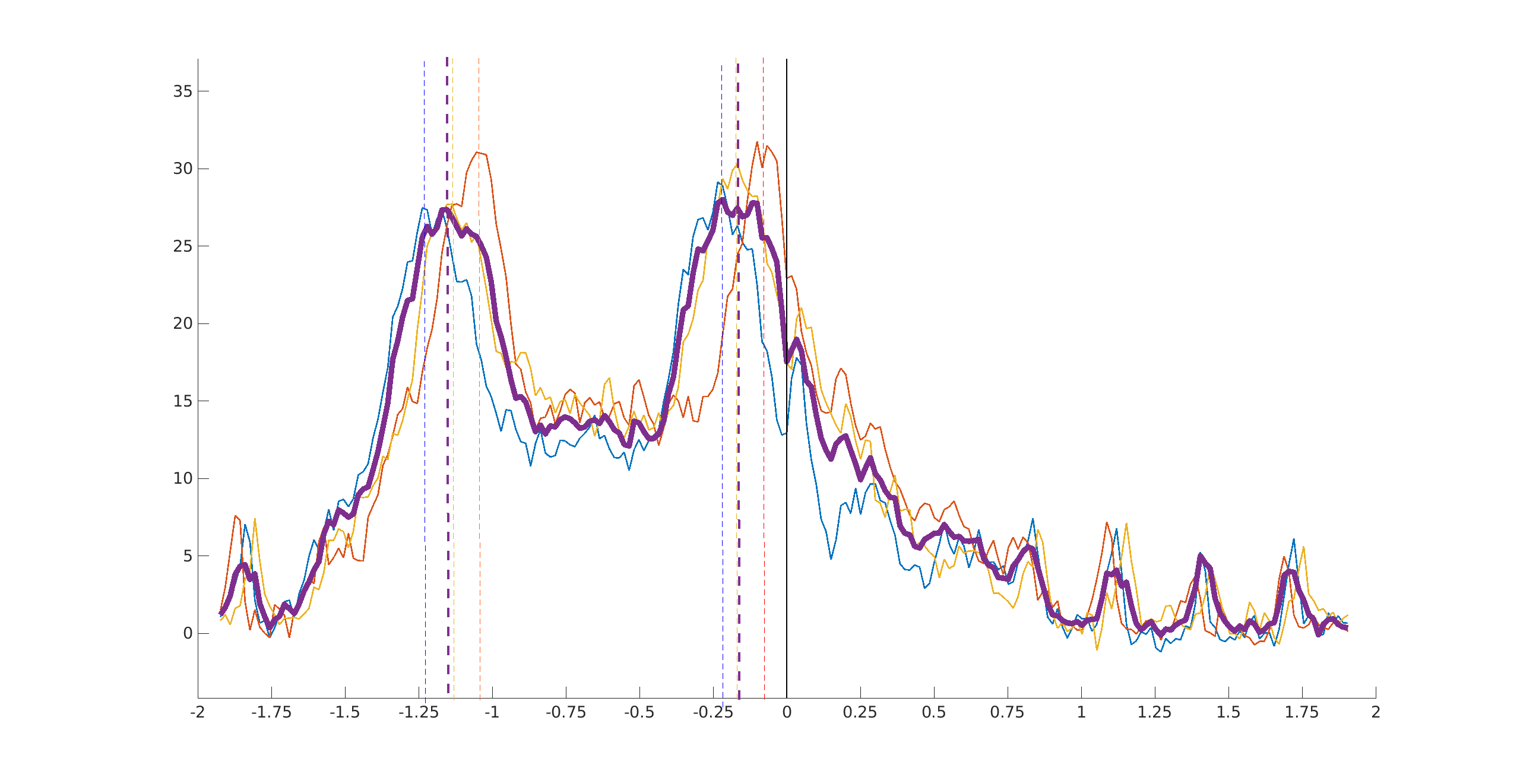}
\caption{\label{HFRResol}{Doppler spectra obtained at the Jobourg station in the high-resolution mode (24.5 MHz, 200 m range resolution) at the range cells 24,27 and 30 in the vicinity of P5 (red, blue, and yellow thin lines, respectively). The thick magenta line shows the Doppler spectrum obtained by averaging the high-resolution spectra between the range cells 24 and 30 to simulate the effect of a coarser resolution of about 1500 m. The dashed vertical lines mark the position of the first-order peaks that would be estimated in each case, showing the important variation of radial current inside the same low-resolution cell.}}
\end{figure}

Another experiment is to compare the surface current radial maps obtained almost simultaneously at the 2 alternate radar frequencies. Figure \ref{fig:2freq} shows successive
radial maps (with a 20 min time delay) obtained with the Jobourg station at 13 MHz and {24.5 MHz} using the 2 types of azimuthal processing (BF or hybrid BF/DF). Note that the available range resolution is coarser at 13 MHz (1500 m) than at 24.5 MHz (750 m). To make them comparable, we have modified the tapering window used for range gating at 13 MHz to increase the effective resolution (at the cost of increased secondary lobes) and bring it closer to the range resolution obtained at {24.5} MHz with the Blackman tapering window. This allows us to isolate the effect of azimuthal resolution alone. Since the same receive antenna array is used at both radar frequencies, the theoretical azimuthal resolution
obtained at 13 MHz is about twice as coarse as the azimuthal resolution obtained at {24.5} MHz ($14^\circ$ versus $7^\circ$). This can be seen quite clearly in Figure \ref{fig:2freq}a), where the fine structures of high current intensity near the coast are smeared when observed at 13 MHz, whereas they are well rendered at {24.5 MHz}. Note that the DF processing (in its improved hybrid version) provides a better description of the sharp fronts that are visible around the spot of the most intense current, regardless of the radar frequency. This shows that with a smaller phased-array (here equivalent to 8 antennas), the DF processing method remains high-performing and can advantageously replace a longer antenna array (16 antennas).

\begin{figure*}[!h]
\begin{minipage}[c]{0.5\textwidth}
\includegraphics[scale=0.5]{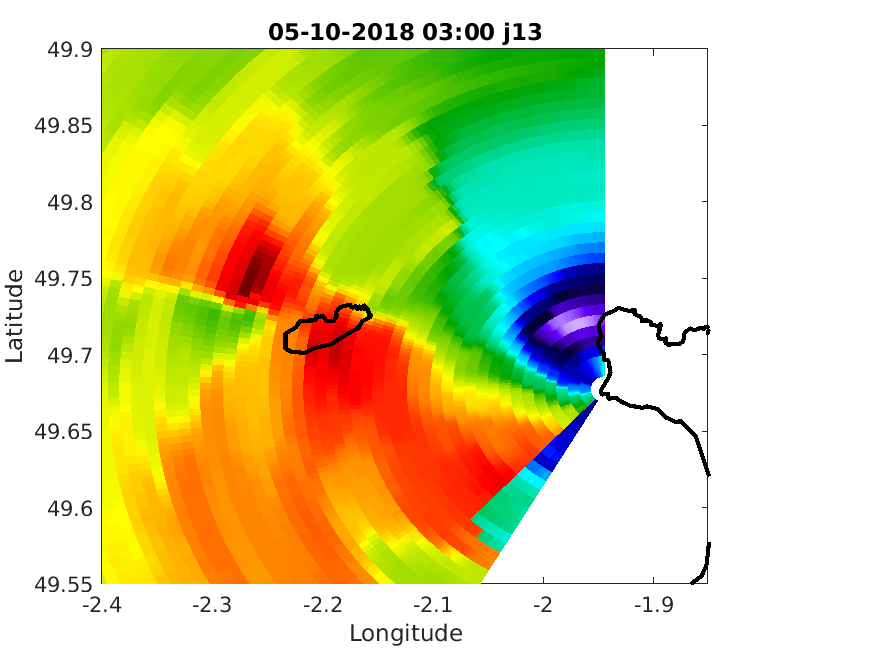}
\end{minipage}\begin{minipage}[c]{0.5\textwidth}
\includegraphics[scale=0.5]{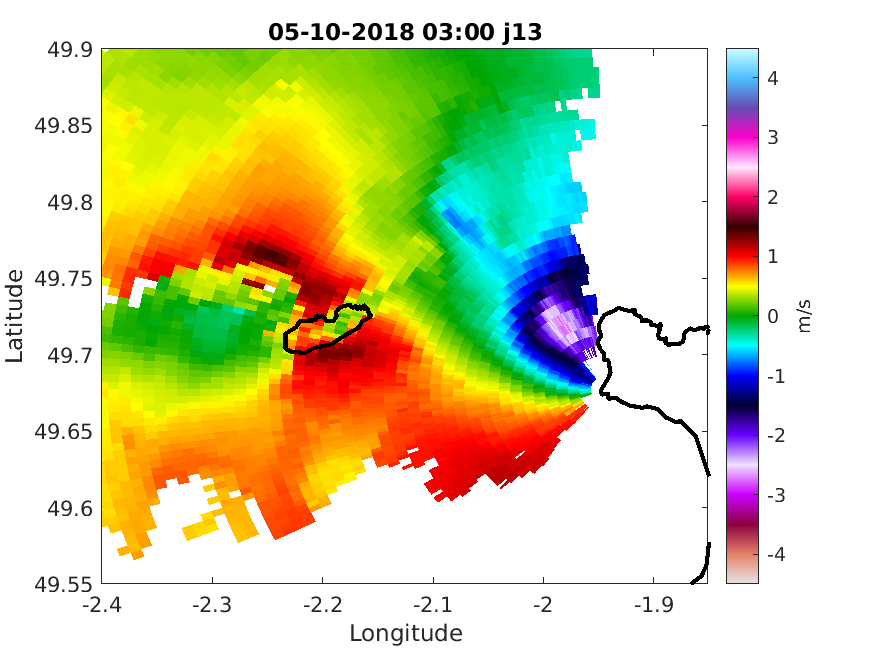}
\end{minipage}
\begin{minipage}[c]{0.5\textwidth}
\includegraphics[scale=0.5]{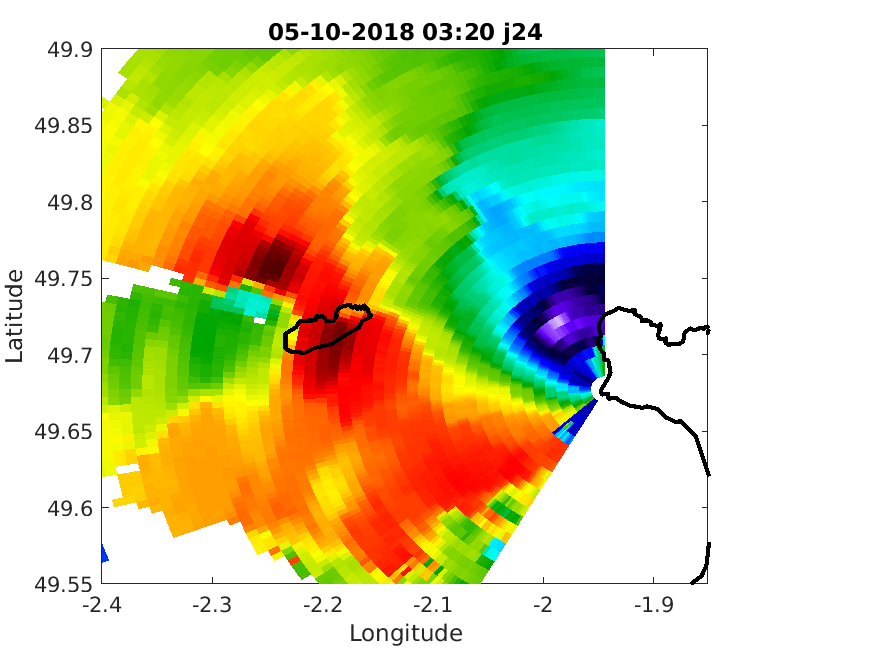}
\end{minipage}\begin{minipage}[c]{0.5\textwidth}
\includegraphics[scale=0.5]{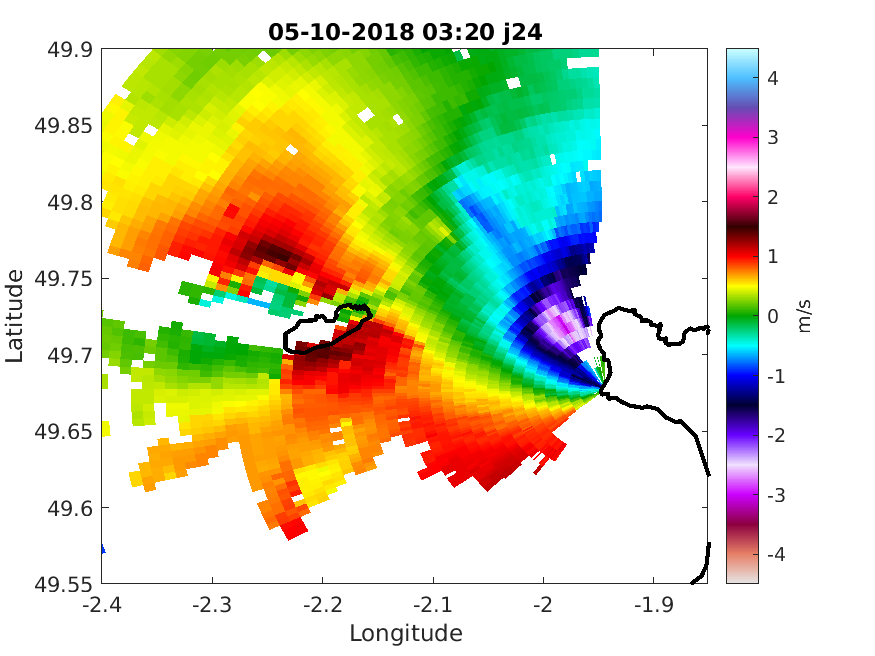}
\end{minipage}
\caption{{Quasi-simultaneous radial maps obtained at the Jobourg station with different radar frequencies and different azimuthal processing: a) 13 MHz BF; b) 13 MHz DF; c) 24.5 MHz BF d) 24.5 MHz DF}. The effective range resolution at the two radar
frequencies is about 1500 m.}
\label{fig:2freq}
\end{figure*}

\section{Conclusion}\label{sec5}

  Strong and sheared currents such as those encountered in the Alderney Race are challenging to measure with HFR. Capturing their spatial heterogeneity requires spatial resolution, which can only be achieved with a wide frequency bandwidth and a large number of antennas. Second, the large Doppler shifts caused by the surface current can cause the Bragg lines to cross the zero Doppler axis, making their identification ambiguous. Finally, in the presence of strong waves and currents, the first- and second-order spectra can overlap in the omnidirectional Doppler spectrum, causing the DF approach to fail.
  We have addressed these issues in this paper. An improved hybrid DF based on BF preconditioning is proposed that eliminates the first/second order ambiguity and, as a by-product, also eliminates the RFI and extends the radar coverage.
  A systematic comparison was made with the classical BF processing available for a phased-array radar system. At the highest operating frequency (24.5 MHz), the 16 antenna array provides an azimuthal resolution of the order of 7 degrees at the beam center, which is sufficient to detect fine flow structures in the near and medium range.
    
 
The improved DF method was evaluated with the classical BF and the HFR measurements were tested against a numerical model (CROCO) at several sites with different bathymetry and against \textit{in situ} measurements, namely an ADCP time series.
  
  In the light of these comparisons, the 16 antenna BF and the DF processing were found to be generally very close and both very consistent with models and ADCP. However, there are a few specific locations with complex and shallow bathymetry where the spatial variations of the current can be very sharp in both space and time, especially during ebb and flow. In this case, the time series of the HFR-based currents show structured asymmetric patterns that the model cannot reproduce, and peaky extrema that have a larger amplitude when estimated with the DF processing.
  We took advantage of an exceptional permit to operate the HFR in a very high spatial resolution mode (200 m) to observe the surface at a very fine scale and to investigate the limits of coarser resolution (750 or 1500 m) in the context of strong and sheared currents. The Doppler spectra measured at high resolution show first-order Bragg lines that are better defined (i.e. thinner) than at low resolution, where they show broader and more fragmented peaks (Fig. \ref{HFRResol}), which can lead to ambiguous estimates of the radial current depending on the technique used. Finally, HFR measurements at 13 and 24.5 MHz were compared, where the theoretical azimuthal resolution using BF processing varies by a factor of two. While BF processing is inaccurate at the lowest radar frequency, the improved DF technique performs equally well in both situations.
  

\bmhead{Acknowledgements}The HF radars were bought by Caen University as a part of the ‘Manche 2021’ project, co-funded by the EU and the Conseil Régional de Normandie, throughout the FEDER-FSE operational programme 2014–2020. The numerical results benefited from CRIANN computing resources. The authors also express their gratitude to L. Perez and M. Boutet for their technical assistance and L. Furgerot for providing the \textit{in situ} data acquired in the framework of HYD2M project funded by the Agence Française pour la Recherche  (ANR-10-IEED-0006-07).
\bmhead{Declarations}
\bmhead{Ethical Approval} Not applicable
 \bmhead{Funding}  Not applicable



\end{document}